\documentclass[a4paper,11pt]{article}
\usepackage{pos}
\usepackage{float}

\title{The Sign-Switching of the Cosmological Constant}
\author*[a,b,c]{Mariam Bouhmadi-López}
\author[b,c]{Beñat Ibarra-Uriondo}

\affiliation[a]{IKERBASQUE, Basque Foundation for Science, 48011, Bilbao, Spain}
\affiliation[b]{Department of Physics, University of the Basque Country UPV/EHU, P.O. Box 644, 48080 Bilbao, Spain}
\affiliation[c]{EHU Quantum Center, University of the Basque Country UPV/EHU, P.O. Box 644, 48080 Bilbao, Spain}

\emailAdd{mariam.bouhmadi@ehu.eus}
\emailAdd{benat.ibarra@ehu.eus}

\abstract{
We propose and investigate a class of dynamical dark energy models in which the cosmological constant evolves from negative values in the early Universe to a positive value at low redshifts. This framework includes a generalised ladder-step evolution, as well as smooth-transition scenarios, providing a unified description of sign-changing cosmological constants. We analyse the theoretical construction and background dynamics of these models using cosmographic diagnostics. Extending this study to the linear perturbation regime, we solve the perturbation equations from the radiation-dominated era with adiabatic initial conditions. We examine the evolution of the matter density contrast, gravitational potential, and the $f\sigma_8$ observable. Our results are compared against the standard $\Lambda$CDM model and confronted with current observational data, illustrating the phenomenological viability of sign-changing dark energy models and revealing distinctive imprints on cosmic structure formation arising from the transition of the cosmological constant.
}

\FullConference{Proceedings of the Corfu Summer Institute 2025 "School and Workshops on Elementary Particle Physics and Gravity" (CORFU2025) 2 - 8 September, 2025
Corfu, Greece
}

\begin{document}
\maketitle

\section{Introduction}
The late-time accelerated expansion of the Universe, first observed through Type Ia supernovae \cite{SupernovaSearchTeam:1998fmf}, is well described by the concordance $\Lambda$CDM model. However, persistent tensions in key cosmological parameters, most notably the $H_0$ and $S_8$ discrepancies \cite{DiValentino:2021izs,Poulin:2018,Kamionkowski:2022pkx,Perivolaropoulos:2021jda,Abdalla:2022yfr}, motivate the exploration of minimal extensions to the standard framework.

A wide range of theoretical extensions beyond the standard cosmological model have been proposed to address the current observational tensions, aiming to reconcile data while preserving the well-established successes of $\Lambda$CDM in describing both early- and late-time cosmic evolution \cite{Bamba:2012cp,CosmoVerse:2025txj}. Broadly speaking, these proposals can be organised into two main directions.

The first focuses on alternative dark energy (DE) frameworks in which the cosmological constant is replaced by a dynamical component. Within this class, one can distinguish between early-time and late-time modifications. Early-time scenarios include, for instance, Early Dark Energy (EDE) \cite{Poulin:2018}, Anti–de Sitter Early Dark Energy (AdS-EDE) \cite{Ye:2020btb,Ye:2020oix,Ye:2021iwa}, and New Early Dark Energy (NEDE) \cite{Niedermann:2019olb,Cruz:2023lmn,Niedermann:2023ssr}. Late-time realisations instead modify the recent expansion history, often through changes in the effective DE density or equation of state (EoS). Representative examples are phantom-crossing DE models \cite{Bouhmadi-Lopez:2008ukq,DiValentino:2020naf,Alestas:2020mvb}, the Omnipotent DE scenario \cite{DiValentino:2020naf,Adil:2023exv}, (non-minimally) interacting DE models \cite{Morais:2016bev,Kumar:2017dnp,Nunes:2022bhn}, and axion-like DE constructions \cite{Kamionkowski:2014zda,Emami:2016mrt,Chiang:2025qxg}.

The second major avenue involves modifications of general relativity itself. In this approach, cosmic acceleration is not attributed to an additional energy component but rather to departures from Einstein’s theory on cosmological scales. Well-studied examples include $f(\mathcal{R})$ gravity \cite{Sotiriou:2008rp,DeFelice:2010aj,Nojiri:2010wj,Nojiri:2017ncd}, $f(\mathcal{T})$ gravity \cite{Bengochea:2008gz,Ferraro:2006jd,Cai:2015emx,Bouhmadi-Lopez:2026dte}, $f(\mathcal{R},T)$ models \cite{Harko:2011kv,Nojiri:2004bi}, and $f(\mathcal{Q})$ theories \cite{BeltranJimenez:2018vdo,BeltranJimenez:2019tme,Ayuso:2020dcu,Boiza:2025xpn,Ayuso:2025vkc}. Additional frameworks include bimetric gravity \cite{Kobayashi:2019hrl} and kinetic gravity braiding (KGB) scenarios \cite{Deffayet:2010qz,Pujolas:2011he,BorislavovVasilev:2022gpp,BorislavovVasilev:2024loq}, among many others.

From the first type we have a particularly promising alternative, the $\Lambda_{\rm s}$CDM model \cite{Akarsu:2021fol,Akarsu:2022typ,Akarsu:2023mfb,Paraskevas:2024ytz,Akarsu:2024qsi,Akarsu:2024eoo,Akarsu:2024nas,Souza:2024qwd,Akarsu:2025gwi,Akarsu:2025ijk,Escamilla:2025imi,Akarsu:2025dmj,Akarsu:2025nns,Ghafari:2025eql,DiGennaro:2022ykp,Yadav:2025vpx}, inspired by the graduated dark energy scenario \cite{Akarsu:2019hmw}. In this framework, the cosmological constant is replaced by an effective dark energy density that undergoes a transition from negative (AdS-like) to positive (dS-like) values at low redshift, $z=z_\dagger \lesssim 2$. This late-time modification leaves early-Universe physics unaffected while significantly alleviating the $H_0$ and $S_8$ tensions \cite{Akarsu:2023mfb,Escamilla:2025imi}. Interestingly, the presence of a negative cosmological constant is theoretically appealing, as AdS vacua naturally arise in string-theoretic and holographic settings \cite{Maldacena:1997re}.

The simplest realisation of $\Lambda_{\rm s}$CDM assumes an abrupt sign switch, typically modelled by a signum function. While phenomenologically successful, this instantaneous transition introduces a discontinuity in the cosmological constant at $z=z_\dagger$, leading to a type~II (sudden) singularity \cite{Nojiri:2005sx}. 
Although such a singularity has been shown to have a negligible impact on the evolution of cosmic structures \cite{Paraskevas:2024ytz}, it nevertheless indicates that the abrupt model should be regarded as an idealised approximation to a more realistic scenario in which the transition is very rapid but smooth, as in graduated dark energy \cite{Akarsu:2019hmw}, while also implementing a finite DE component in the past.

%Although such a singularity has been shown to have a negligible impact on the evolution of cosmic structures \cite{Paraskevas:2024ytz}, it indicates that the abrupt model should be regarded as an idealised approximation of a more realistic scenario where transition is rapid rapid but smooth like in graduated dark energy however implenting as well a finite DE in the past. {\color{red} keeping as well a non-divergent behaviour at early-times unlike graduated dark energy. To be discussed with Beñat.}{\color{blue} keeping as well a bounded behaviour at early times unlike gradated dark energy.....Perhaps more subtle?} {\color{red} puedes redactar lo frase, no entiendo muy bien lo que quieres decir.}.

Motivated by this, we have proposed several smooth extensions of the abrupt $\Lambda_{\rm s}$CDM model \cite{Bouhmadi-Lopez:2025ggl} (see also \cite{Akarsu:2024qsi,Akarsu:2024eoo,Souza:2024qwd,Akarsu:2024nas} for alternative smooth behaviours). These include models featuring a ladder-like cosmological constant, a smooth step function, and an error-function profile. By construction, these scenarios eliminate sudden singularities, replacing them with milder $w$-singularities, while preserving the successful background evolution of the original model. Importantly, these smooth transitions allow for a consistent perturbative treatment, enabling the study of structure formation through linear perturbation theory.

The impact of these models on the growth of matter perturbations can be directly tested using large-scale structure observables such as the growth rate $f\sigma_8$ and the matter power spectrum. Preliminary analyses indicate that smooth sign-switching models remain compatible with current observational data while providing a more physically motivated description of late-time DE dynamics.

This manuscript is based on our  work \cite{Bouhmadi-Lopez:2025ggl,Bouhmadi-Lopez:2025spo} and is organised as follows. In Sec.~\ref{sec2} we introduce and compare the abrupt sign-switching model with its three smoother extensions. In Sec.~\ref{sec3} we present the evolution of the cosmographic parameters of these models. In Sec.~\ref{sec4} we examine the evolution of linear matter perturbations, the gravitational potential, and structure formation using several measurements of the growth rate function, specifically $f\sigma_8$. We conclude briefly in Sec.~\ref{sec5}.

\section{Beyond the standard model\label{sec2}}

In this section, we introduce and compare the abrupt sign-switching model with three smoother extensions. We present the phenomenological evolution of each dark energy (DE) density in turn. Our analysis is carried out within the standard framework of a homogeneous and isotropic cosmological background described by the Friedmann–Lemaître–Robertson–Walker (FLRW) line element. We restrict attention to a spatially flat geometry and model the cosmic energy content as a set of non-interacting components: radiation, pressureless matter — comprising both baryonic and cold dark matter contributions — and DE.

\subsection{Abrupt sign-switching model (\texorpdfstring{$\Lambda_{\rm s}$}{Lambda{s}}CDM) -- Model (A) \label{sec2-A}}

We start by considering the most elementary  realisation of a sign-changing DE model, namely $\Lambda_{\rm s}$CDM \cite{Akarsu:2021fol}. This single-parameter extension of $\Lambda$CDM is characterised by a DE density that transitions from a constant negative value to a constant positive one via a step function, namely,
\begin{equation}
\Lambda_\mathrm{s}(z) = \Lambda \text{ sgn}(z_{\dagger,s} - z),
\end{equation}
where $z_{\dagger,s}$ denotes the redshift at which the sign-switching occurs, and $\Lambda$ represents the 
present-day value of the cosmological constant, at which the model reduces to the well-known
$\Lambda$CDM scenario. The function $\mathrm{sgn}(z_{\dagger,s} - z)$ represents the signum function, taking the value $1$ when its argument is positive and $-1$ when it is negative. The EoS parameter remains constant at all times, mimicking $\Lambda$CDM except at the sign-switching, when it diverges,
\begin{equation}
    w_{\textrm{d,s}}=-1-\frac{\updelta(z_{\dagger,s}-z)}{3 \textrm{ sgn}(z_{\dagger,s}-z)}(1+z).
\end{equation}
Here $\updelta(z_{\dagger,s}-z)$ stands for the Dirac delta function. It follows that the EoS parameter becomes undefined at the moment of the sign change, giving rise to a singular behaviour that was analysed in \cite{Paraskevas:2024ytz}. Additional details of this model can be found in Ref. \cite{Akarsu:2021fol}.

\subsection{Ladder-like DE model (L\texorpdfstring{$\Lambda$}{Lambda}CDM) -- Model (B)\label{sec2-B}}

This framework represents a natural generalisation of the previous case, extending from a single step to $N$ successive steps. Each step introduces a discrete increase in the DE density via a ''small'' discontinuous function. The size of each step is determined by the total number of steps, and for simplicity, all steps are taken to be equal. Throughout this work, we fix $N=20$, thereby rendering the model effectively a two-parameter extension of $\Lambda$CDM, defined by an initial and final redshift, $z_{i,le}$ and $z_{f,le}$, which mark the beginning and end of the transition from $-\Lambda$ to $+\Lambda$. The growth of the cosmological constant is encoded through an $N$-step ladder function, with the length of each step serving as a controllable parameter
\begin{equation}
  \begin{aligned}
&  \Lambda_\mathrm{l}(z)=\Lambda \left[1-\frac{2}{N}\sum_{n=1}^{N} \mathcal{H}(z_n-z)\right].
\end{aligned}  
\label{densityladder}
\end{equation}

The redshift at which each step transition occurs is given by
\begin{equation}
\begin{aligned}
    z_n = & z_{f,l} + \frac{n}{N} \left( z_{i,l} - z_{f,l} \right), \\
    z_{i,l}= & z_{\dagger,l}+\Delta z\cdot \frac{N}{2}, \\
    z_{f,l}= & z_{\dagger,l}-\Delta z\cdot \frac{N}{2},
\end{aligned}
\label{setpz}
\end{equation}
where $n \in [0,N]$ indexes the steps, ordered from the present to the past, $z_{\dagger,l}$ is the sign-switching redshift and $\Delta z$ the length of each step. The function $\mathcal{H}(z_n - z)$ denotes the Heaviside step function evaluated at $z_n - z$. As in the simpler single-step case, the EoS parameter exhibits singularities at each discontinuity corresponding to the step transitions:
 \begin{equation}
    w_{\textrm{d,l}}=-1-\frac{(1+z)\sum^{N}_{i=1}\updelta(z_n-z)}{3 \left[-N/2+\sum^{N}_{n=1}\mathcal{H}(z_n-z)\right]}.
    \label{eosl}
\end{equation}

In this model, the EoS parameter remains $-1$ throughout all regimes, except at the locations of the step transitions, where it diverges owing to the presence of Dirac delta functions.

\subsection{SSCDM -- Model (C)\label{sec2d}}

The third model represents another two-parameter extension of $\Lambda$CDM, in which the DE density evolves smoothly from negative to positive values over a specified redshift range through an interpolation polynomial. The additional parameters are the redshift at which the transition commences, $z_{i,ss}$, and the redshift at which it concludes, $z_{f,ss}$; these parameters control the smoothness of the transition. Within this framework, the DE density is expressed as:

\begin{equation}
  \begin{aligned}
&  \Lambda_\mathrm{ss}(x)=\Lambda \begin{cases}-1, & x \le x_{i,ss},\\  1-2(126 t^5 - 420 t^6 + 540 t^7 - 315 t^8 + 70 t^9),  & x_{i,ss}<x<x_{f,ss},\\
1, & x\ge x_{f,ss},\end{cases} \\
\end{aligned}  
\label{dnesitysscdm}
\end{equation} 

where
\[
t = \frac{x - x_{f,ss}}{x_{i,ss} - x_{f,ss}}, \qquad x = -\ln(1+z).
\]
In these class of models, the EoS parameter departs from the $\Lambda$CDM behaviour, since the DE density is no longer constant over part of the cosmic evolution. Consequently, the EoS parameter becomes time-dependent and evolves as

\begin{equation}
  \begin{aligned}
&  w_\mathrm{d,ss}(x)= \begin{cases}-1, & x \le x_{i,ss},\\-1  -\frac{1260}{\Delta x \text{ }\Lambda_\mathrm{ss}(x)/\Lambda}(t^4 -4t^5+6t^6-4t^7+t^8),  & x_{i,ss}<x<x_{f,ss},\\
-1, & x\ge x_{f,ss},\end{cases}\\
\end{aligned}  
\label{eossscdm}
\end{equation}
 where $\Delta x = x_{i,ss} - x_{f,ss}$. In models (A) and (B), the EoS parameter contains Dirac delta contributions at each step discontinuity, originating from the derivative of the step function. This behaviour is absent in the present model, as the transition in the DE density is implemented in a smooth manner. 

Nevertheless, as is generic in DE models featuring a transition from negative to positive energy density, the EoS parameter $w_{\mathrm{d,ss}}(x)$ diverges at the point where the sign change occurs, undergoing a crossing of the cosmological constant line, $w=-1$, through infinite values. Although $w_{\mathrm{d,ss}}(x)$ is not well defined over the entire interval $x \in (x_{i,ss}, x_{f,ss})$, the total equation-of-state parameter, $w_{\rm tot} = \Omega_{\rm r}/3 + w_{\rm d}\Omega_{\rm d}$, where $\Omega_{\rm r}$ and $\Omega_{\rm d}$ denote the radiation and DE fractional energy densities, respectively, remains well behaved throughout the evolution.

\subsection{ECDM -- Model (D)\label{sec2e}}
 In this model, we introduce a DE density that evolves according to an interpolating error function\footnote{The error function is a function erf: $\mathbb{C}\rightarrow\mathbb{C}$ defined as: $$\operatorname{erf}(z)=\frac{2}{\sqrt{\pi}} \int_0^z e^{-t^2} d t\, .$$}. Specifically, we assume
\begin{equation}
    \Lambda_\mathrm{e}(x) = \Lambda\, \mathrm{erf}\!\left[\eta \left(x - x_{\dagger,e}\right)\right],
    \label{densityecdm}
\end{equation}
where $\eta$ is a parameter controlling the smoothness of the transition, and $x_{\dagger,e} = -\ln(1 + z_{\dagger,e})$ denotes the redshift at which the DE density changes sign. This construction therefore constitutes a two-parameter extension of the standard $\Lambda$CDM framework.

The corresponding EoS parameter can be straightforwardly derived by enforcing conservation of the DE energy--momentum tensor, yielding
\begin{equation}
    w_\mathrm{d,e}(x) = -1 - \frac{2\eta\, \mathrm{e}^{-\eta^2 (x - x_{\dagger,e})^2}}{3\sqrt{\pi}\,\mathrm{erf}\!\left[\eta \left(x - x_{\dagger,e}\right)\right]}.
    \label{eosecdm}
\end{equation}

As in model (C), the smooth evolution of the DE density in this scenario removes the Dirac delta contributions to the EoS parameter that are present in models (A) and (B). Nonetheless, similarly to the previous case, the vanishing of the error function at the sign-changing point $x = x_{\dagger,e}$ causes the EoS parameter to diverge, resulting in a cosmological constant line crossing through infinite values. By contrast, the total EoS parameter, $w_{\rm tot}=\Omega_{ \rm r}/3+w_{\rm d}\Omega_{\rm d}$, remains well defined, exhibiting a regular background evolution throughout.

 \subsection{Model comparison}

 A brief comparison of the different models, and in particular with $\Lambda$CDM, is now in order. At early times, all models exhibit identical behaviour; the same is true once the cosmological constant has attained its final constant value, at which point they all effectively reduce to $\Lambda$CDM. Distinctive features arise only during the transition phase and are primarily associated with the manner and smoothness of the evolution.

 In order to perform a meaningful comparison among the models, it is necessary to specify the parameters that characterise each of them. The procedure adopted for this purpose is outlined as follows. We begin by specifying the general quantities employed in all numerical simulations. Following Ref.~\cite{Akarsu:2022typ} and adopting \textit{Planck 2018} data, we fix the $\Lambda$CDM parameters to
$H_{0,\Lambda \mathrm{CDM}} = 67.29~\mathrm{km\,s^{-1}\,Mpc^{-1}}$,
$\Omega_{\mathrm{m0},\Lambda \mathrm{CDM}} = 0.3125$,
$\Omega_{\mathrm{r0},\Lambda \mathrm{CDM}} = 9 \times 10^{-5}$,
and
$\Omega_{\rm d,\Lambda \mathrm{CDM}} = 1 - \Omega_{\mathrm{m0},\Lambda \mathrm{CDM}} - \Omega_{\mathrm{r0},\Lambda \mathrm{CDM}}$.
For the sign-switching models, we instead adopt the parameter values obtained in the same work for the abrupt sign-switching scenario, denoted as $\Lambda_s$CDM, namely
$H_0 = 70.22~\mathrm{km\,s^{-1}\,Mpc^{-1}}$,
$\Omega_{\mathrm{m0}} = 0.2900$,
$\Omega_{\mathrm{r0}} = 9 \times 10^{-5}$,
and
$\Omega_{\mathrm{d0}} = 1 - \Omega_{\mathrm{m0}} - \Omega_{\mathrm{r0}}$.

Turning to the characteristic parameters of each model, we fix the sign-switching redshift of model (A) to $z_{\dagger,s} = 1.7$, in agreement with Refs.~\cite{Akarsu:2021fol,Akarsu:2022typ,Akarsu:2023mfb,Escamilla:2025imi}. For model (B), we adopt the same sign-switching redshift, $z_{\dagger,l} = 1.7$, to enable a direct comparison. Assuming a rapid transition, we set the step width to $\Delta z = 0.1$, with the transition beginning at $z_{i,l} = 2.7$ and ending at $z_{f,l} = 0.7$, in accordance with Eq.~(\ref{setpz}). In model (C), the additional parameters are chosen such that the sign-switching occurs at a comparable redshift and over a transition interval of similar duration. Finally, for model (D), we fix the sign-switching redshift to the same value as in models (A) and (B), while selecting a transition speed comparable to that of model (C) and to the previously studied $\Lambda_s$VCDM scenario \cite{Akarsu:2024eoo}. For clarity, we summarise the parameter choices as follows: model (A) is characterised by $z_{\dagger,s} = 1.7$; model (B) by $z_{\dagger,l} = 1.7$, and $\Delta z = 0.1$; model (C) by $z_{i,ss} = 2.7$ and $z_{f,ss} = 1$; and model (D) by $\eta = 10$ and $z_{\dagger,e} = 1.7$.

\begin{figure}[htbp]
    \centering
    
    % Top row: two figures side by side
    \begin{tabular}{cc}
        \includegraphics[width=0.48\linewidth]{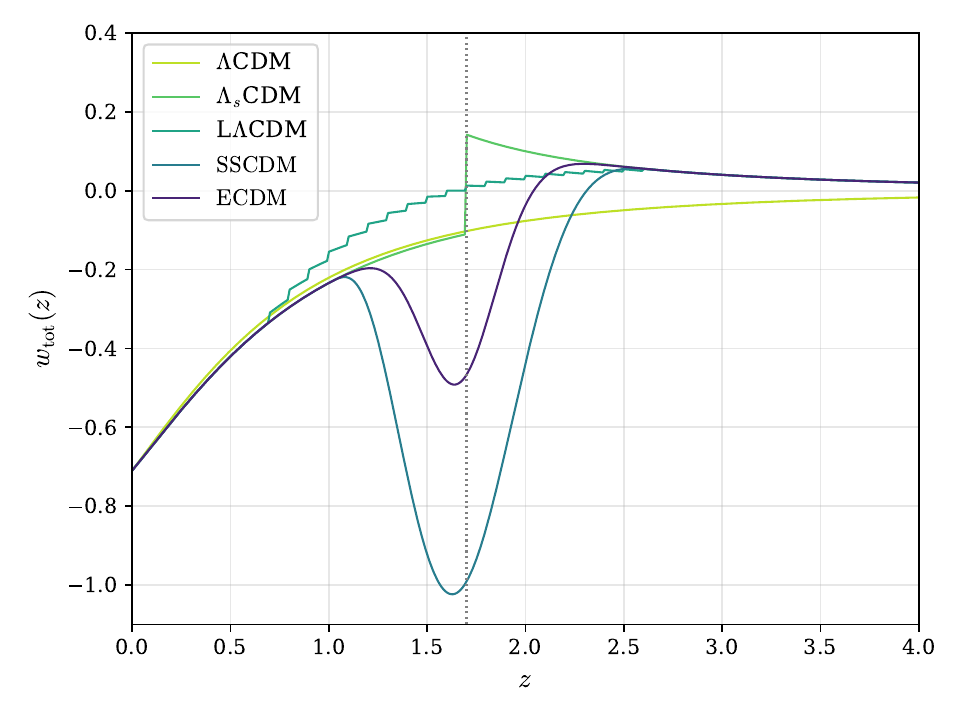} & 
        \includegraphics[width=0.48\linewidth]{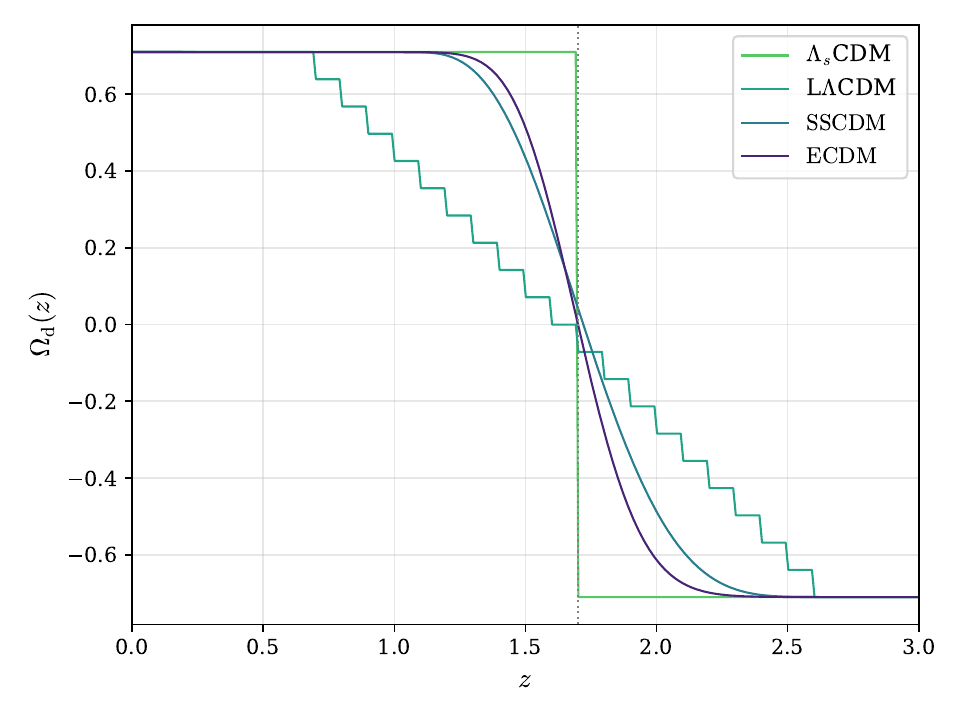} \\
    \end{tabular}
    
    % Vertical space between rows
    \vspace{0.5cm}
    
    % Bottom row: single centered figure
    \begin{tabular}{c}
        \includegraphics[width=0.48\linewidth]{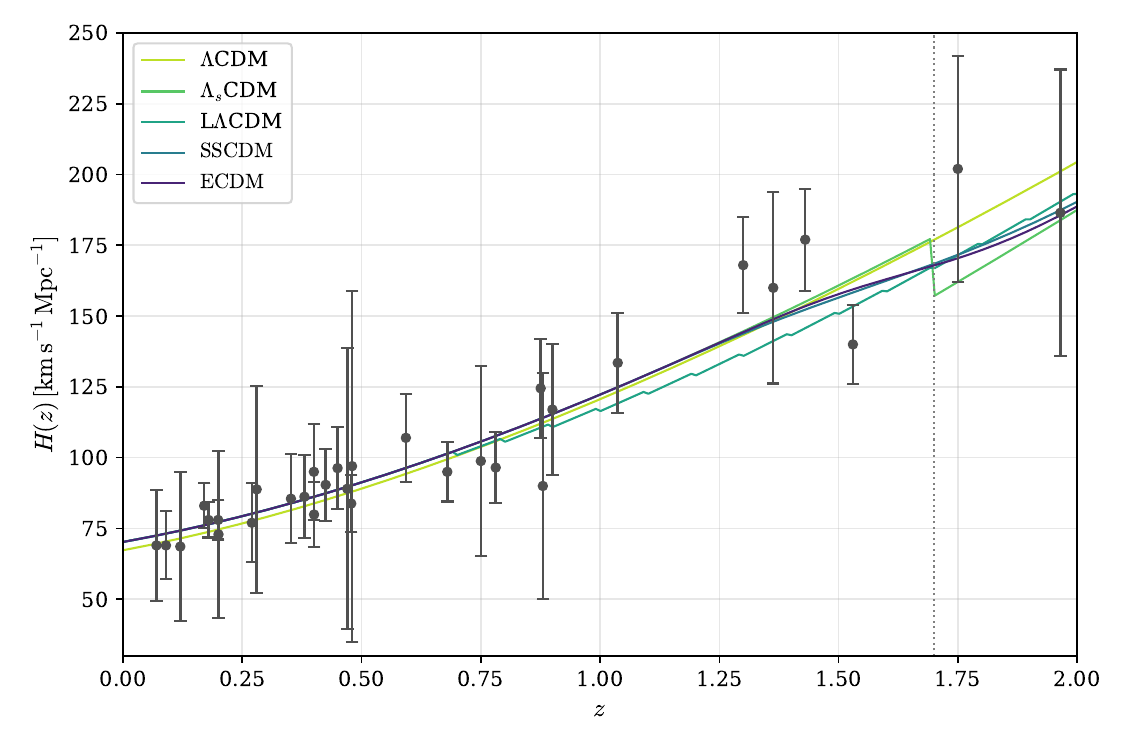} \\
    \end{tabular}
    
    \caption{The left figure showcases the total EoS parameter, $w_{\mathrm{tot}}(z)$, against redshift. The right figure showcases the evolution of DE energy density. The bottom figures showcases $H(z)$ in units of $km/s/Mpc$ against redshift.  The the error bars were obtained from Ref.~\cite{Favale:2023lnp}. The vertical black line denotes the sign-switching redshift for the DE density. }
    \label{fig:hubble}
\end{figure}

These models can be naturally divided into two distinct classes. The first class comprises models (A) and (B), which display discontinuous behaviour across their defining features. This is evident in both the Hubble parameter $H(z)$ and the EoS parameter, as illustrated in Fig.~\ref{fig:hubble}. While the piecewise-constant nature of the DE density simplifies the description of several quantities—since the EoS parameter remains fixed and DE perturbations are absent—the abrupt variations in the DE density give rise to sudden-type singularities \cite{Paraskevas:2024ytz, Bouhmadi-Lopez:2025ggl}. The second class includes models (C) and (D), which instead exhibit fully continuous dynamics. In these scenarios, the Hubble parameter evolves smoothly and the EoS parameter becomes time-dependent. Such variations are expected to play a more prominent role in the cosmographic analysis presented in the following section.

When introducing the models, we noted that all of them exhibit a divergence at the sign-switching point, which arises from the vanishing of the DE density at that instant. Since the EoS parameter is defined as $w_{\mathrm{d}} = p_{\mathrm{d}} / \rho_{\mathrm{d}}$, it necessarily diverges when $\rho_{\mathrm{d}} \to 0$. This feature is common to all sign-switching models. Nevertheless, as shown in Fig.~\ref{fig:hubble}, the total equation-of-state parameter remains well defined throughout the entire evolution. For all sign-switching scenarios, the total EoS parameter exceeds the $\Lambda$CDM value prior to the sign change and becomes slightly smaller afterwards. In particular, for the two continuous models, an oscillatory behaviour can be observed, which is expected to become more pronounced in the cosmographic analysis presented in the following section.
 
\section{Cosmography\label{sec3}}

A convenient way to contrast the models discussed above is through a cosmographic analysis \cite{Visser:2004bf,Cattoen:2007sk,Capozziello:2008qc}. This approach is particularly well suited for comparative studies, as it allows one to characterise different cosmological scenarios in terms of a common set of observable parameters.

Within this framework, the scale factor is expanded as a Taylor series around its present-day value $a_0$, namely
\begin{equation}
    a(t) = a_0 \left[ 1 + \sum_{n=1}^{\infty} \frac{A_n(t_0)}{n!} \left[ H_0 (t - t_0) \right]^n \right],
\end{equation}
where $t_0$ denotes the present cosmic time and $H_0$ is the Hubble parameter evaluated today. The coefficients of this expansion define the cosmographic parameters $A_n$, which are given by
\begin{equation}
    A_n \equiv \frac{a^{(n)}}{a H^n}, \qquad n \in \mathbb{N},
    \label{an}
\end{equation}
with $a^{(n)}$ representing the $n$-th derivative of the scale factor with respect to cosmic time.

Several of these parameters have acquired standard names in the literature due to their direct physical interpretation. In particular, the second-order parameter is related to the deceleration parameter, $q = -A_2$, while higher-order terms define the jerk, $j = A_3$, the snap, $s = A_4$, and the lerk, $l = A_5$. These quantities will play a central role in the comparison of the different sign-switching models discussed in this work.

\subsection{Cosmographic parameters}

In this section, we focus on the behaviour of the cosmographic parameters at low redshift. In Ref.~\cite{Bouhmadi-Lopez:2025ggl}, we derived the analytical expressions of these parameters up to the lerk and showed that, for the continuous transition models, higher-order derivatives of the scale factor naturally lead to an increasing number of oscillatory features. This behaviour is expected, as successive time derivatives amplify any underlying variation in the expansion history.
For clarity of presentation, and due to space limitations, the results are displayed in two separate columns. This choice is particularly motivated by the continuous models, whose cosmographic parameters exhibit larger amplitudes and richer structure compared to the discontinuous cases.

\begin{figure}[h]
\centering

\begin{tabular}{cc}
\includegraphics[width=0.45\textwidth]{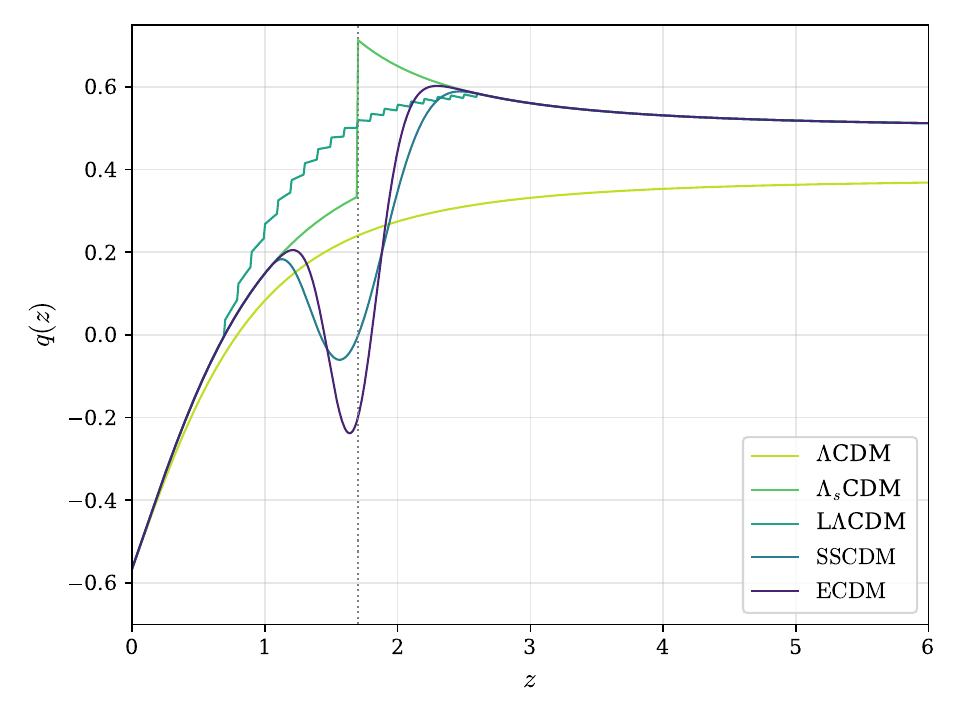} &
\includegraphics[width=0.45\textwidth]{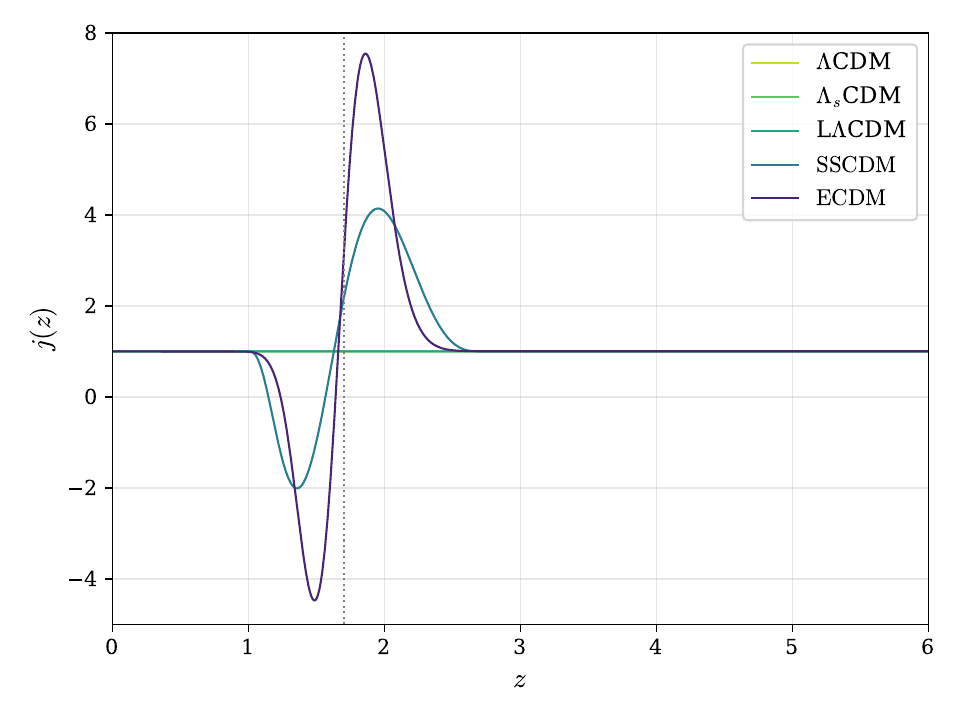} \\[0.5em]

\includegraphics[width=0.45\textwidth]{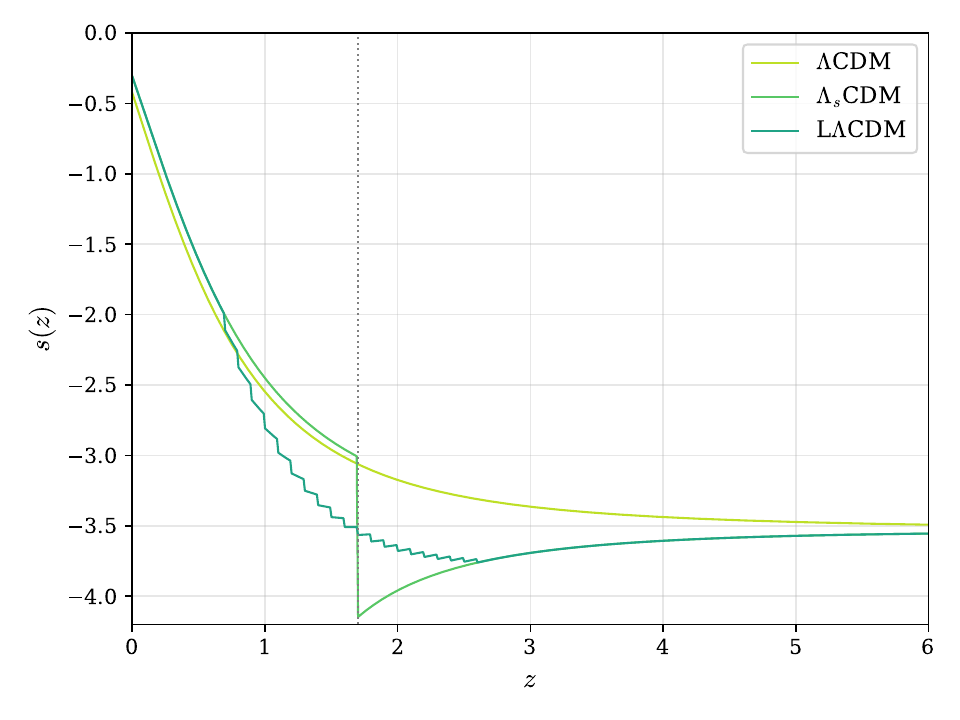} &
\includegraphics[width=0.45\textwidth]{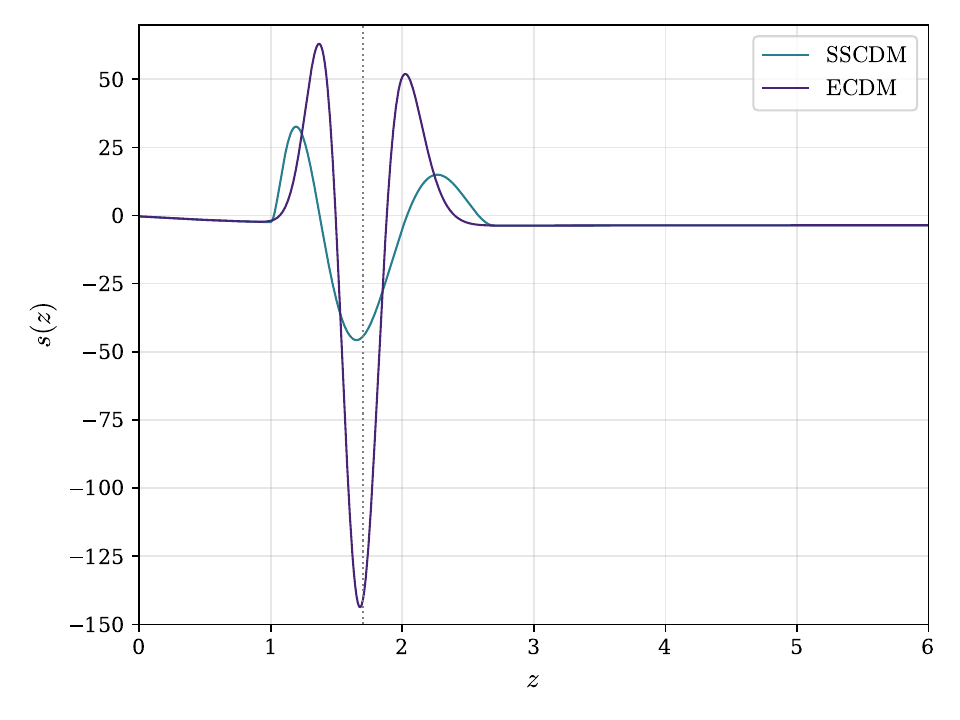} \\[0.5em]

\includegraphics[width=0.45\textwidth]{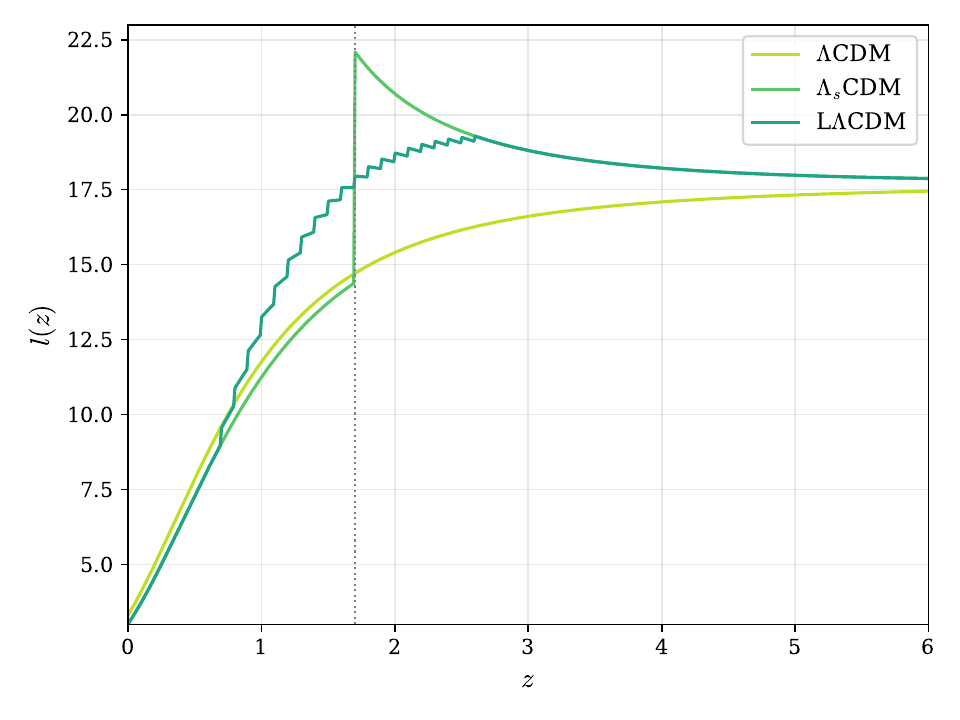} &
\includegraphics[width=0.45\textwidth]{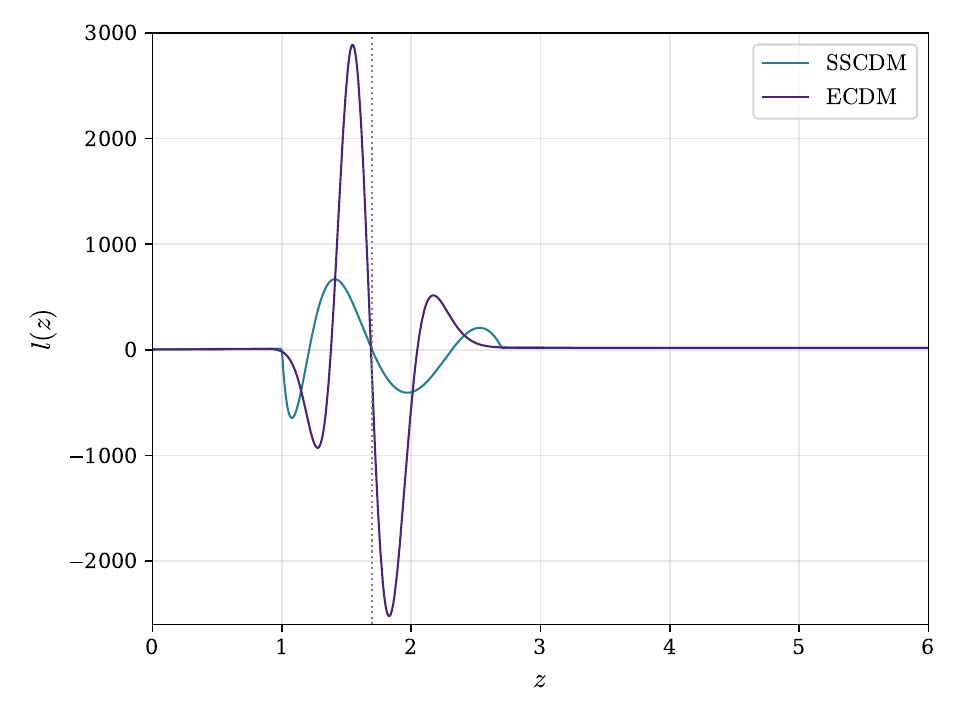}
\end{tabular}

\caption{Comparison of the cosmographic parameters $q(z)$, $j(z)$, $s(z)$ and $l(z)$ for the different models. The vertical line denotes the redshift at which the DE density changes sign. }
\label{fig:all_plots}
\end{figure}

The top-left panel of Fig.~\ref{fig:all_plots} illustrates the redshift evolution of the deceleration parameter, $q(z)$, for all the models considered in this work. In the discontinuous scenarios, namely models (A) and (B), the presence of a negative DE density prior to the sign-switching leads to an effective positive pressure. This results in an enhanced deceleration when compared to the standard $\Lambda$CDM case. Following the sign change, these models enter an accelerated expansion phase at a redshift slightly later than that of $\Lambda$CDM, eventually converging towards the same asymptotic value. The continuous models, (C) and (D), display a qualitatively different behaviour. One of their most distinctive features is the emergence of an additional, intermediate phase of accelerated expansion. As in the discontinuous cases, these models experience a stronger decelerating phase before the transition. However, owing to the rapid yet smooth nature of the transition, the deceleration parameter crosses the threshold $q=0$, signalling the onset of a new accelerated epoch around $z \sim 1.8$, that is, after the sign-switching of the DE density, in agreement with the results from Ref.~\cite{Ibarra-Uriondo:2026zbp,Akarsu:2026anp}. This phase is short-lived: the Universe subsequently returns to a decelerating regime before finally entering the present-day accelerated expansion, where all models ultimately converge.

The top-right panel of Fig.~\ref{fig:all_plots} shows the redshift dependence of the jerk parameter, $j(z)$, for all the models analysed here. In the scenarios with a constant DE density, the jerk is largely determined by the present radiation abundance, $\Omega_{\mathrm{r0}}$, and consequently remains close to the $\Lambda$CDM value, $j(z)=1$, throughout the low-redshift regime. In contrast, the continuous transition models display the oscillatory features already discussed in the previous section. These oscillations become increasingly pronounced for higher-order derivatives, as expected within a cosmographic framework. Nonetheless, once the transition region is left behind, the jerk parameter in these models also relaxes back towards the standard behaviour, asymptotically approaching $j(z)=1$.

The middle row of Fig.~\ref{fig:all_plots} presents the evolution of the snap parameter, $s(z)$, highlighting distinct behaviours across the different classes of models. The left-middle panel corresponds to scenarios with a constant DE density. In these cases, the snap parameter takes values lower than those of $\Lambda$CDM prior to the sign-switching, before converging towards slightly larger values at the present epoch.
The right-middle panel displays the behaviour of the continuous models, (C) and (D), which are characterised by pronounced oscillations in $s(z)$ around the sign-switching redshift. As the DE density gradually settles to its $\Lambda$CDM limit, the snap parameter increases and eventually aligns with the behaviour observed in the discontinuous models (A) and (B).

The bottom row of Fig.~\ref{fig:all_plots} is devoted to the evolution of the lerk parameter, $l(z)$. The left-bottom panel shows the models with a constant DE density. Before the sign-switching occurs, these scenarios predict lerk values that are systematically larger than those of $\Lambda$CDM. After the transition, the trend is reversed, with the models converging towards slightly smaller present-day values. The right-bottom panel displays the continuous transition models, (C) and (D), which again exhibit the characteristic oscillatory behaviour during the sign-switching epoch. These oscillations, common to all cosmographic parameters in the continuous models, may provide a simple and effective diagnostic for testing their phenomenological viability. In particular, the absence of such oscillatory features—should future observations allow for accurate measurements of higher-order cosmographic parameters—would place strong constraints on, or potentially rule out, this class of models.

\section{Cosmological perturbations\label{sec4}}

Starting from the perturbed FLRW metric in the Newtonian gauge we can compute the perturbed Einstein equations \cite{bauman,Brandenberger:1993zc,Ma:1995ey,2004,Malik_2009}. Then  following a multi-fluid approach, as done, for example,  in \cite{Albarran:2016mdu}, and moving to Fourier-space, from the perturbation of the conservation of the energy-momentum tensor we obtain the evolution equation for the fractional energy density perturbation\footnote{$\delta$ denotes a perturbative variable, not to be confused with the Dirac delta function $\updelta(z)$.}, $\delta_i=\delta\rho_i/\rho_i$, and the peculiar velocity potential, $v_i$, for each mode and for the different
components of the universe:
\begin{equation}
\begin{aligned}
 \left(\delta_{\mathrm{r}}\right)_x=&\frac{4}{3}\left(\frac{k^2}{\mathcal{H}} v_{\mathrm{r}}+3 \Psi_x\right), \\
 \left(v_r\right)_x=&-\frac{1}{\mathcal{H}}\left(\frac{1}{4} \delta_{\mathrm{r}}+\Psi\right), \\
\left(\delta_{\mathrm{m}}\right)_x=&\left(\frac{k^2}{\mathcal{H}} v_{\mathrm{r}}+3 \Psi_x\right), \\
 \left(v_{\mathrm{m}}\right)_x=&-\left(v_{\mathrm{m}}+\frac{\Psi}{\mathcal{H}}\right),\\
 \left(\delta_{\mathrm{d}}\right)_x=&\left(1+w_{\mathrm{d}}\right)\left\{\left[\frac{k^2}{\mathcal{H}}+9 \mathcal{H}\left(c_{s \mathrm{~d}}^2-c_{a \mathrm{~d}}^2\right)\right] v_{\mathrm{d}}+3 \Psi_x\right\}+3\left(w_{\mathrm{d}}-c_{s \mathrm{~d}}^2\right) \delta_{\mathrm{d}}, \\
 \left(v_{\mathrm{d}}\right)_x=&-\frac{1}{\mathcal{H}}\left(\frac{c_{s \mathrm{~d}}^2}{1+w_{\mathrm{d}}} \delta_{\mathrm{d}}+\Psi\right)+\left(3 c_{s \mathrm{~d}}^2-1\right) v_{\mathrm{d}},
\end{aligned}
\label{perturb}
\end{equation}
and for the metric potential

\begin{equation}
\begin{aligned}
&\Psi_x+\Psi\left(1+\frac{k^2}{3 \mathcal{H}^2}\right)  =-\frac{1}{2} \delta, \\
&\Psi_x+\Psi  =-\frac{3}{2} \mathcal{H} v(1+w), \\
&\Psi_{x x}+\left[3-\frac{1}{2}(1+3 w)\right] \Psi_x-3 w \Psi  =\frac{3}{2} \frac{\delta p}{\rho},
\end{aligned}
\label{psieq}
\end{equation}
where $\{\}_x$ indicates derivatives with respect to $\ln(a/a_0)$, $\Psi$ denotes the gravitational potential, $k$ is the Fourier wavenumber, and $\mathcal{H}$ is the Hubble parameter expressed in conformal time. For models (A) and (B), perturbations are absent owing to the constant character of their DE density. In the case of Models (C) and (D), DE perturbations are in principle present. However, owing to the assumption of a rapid transition, the DE component remains dynamical only for a very brief interval. Consequently, DE perturbations are expected to remain nearly constant for the majority of the evolution and to exert a negligible influence on the overall dynamics. For this reason, we do not include this component in our analysis.

\subsection{Numerical results}
In this section, we investigate the evolution of the perturbative variables $\delta_{\mathrm{m}}$, $v_{\mathrm{m}}$, $\delta_{\mathrm{r}}$ and $v_{\mathrm{r}}$ by numerically integrating the coupled system of Eqs.~(\ref{perturb}) and (\ref{psieq}). The integration is performed from deep within the radiation-dominated era, corresponding to $x=-15$ ($z \sim 10^{6}$), up to the future epoch $x=2$ ($z=-0.5$). Initial conditions are imposed following the prescription of Ref.~\cite{Bouhmadi-Lopez:2025spo}. The numerical procedure is repeated for a range of Fourier modes spanning from $k = 0.1\, \rm h\,\mathrm{Mpc}^{-1}$ down to $k = 3.33 \times 10^{-4}\,\rm h\,\mathrm{Mpc}^{-1}$, allowing us to probe both small- and large-scale perturbations across cosmic time.

\subsubsection{Matter perturbations}

\begin{figure}[t]
\centering

\begin{tabular}{cc}
\shortstack{\includegraphics[width=0.45\textwidth]{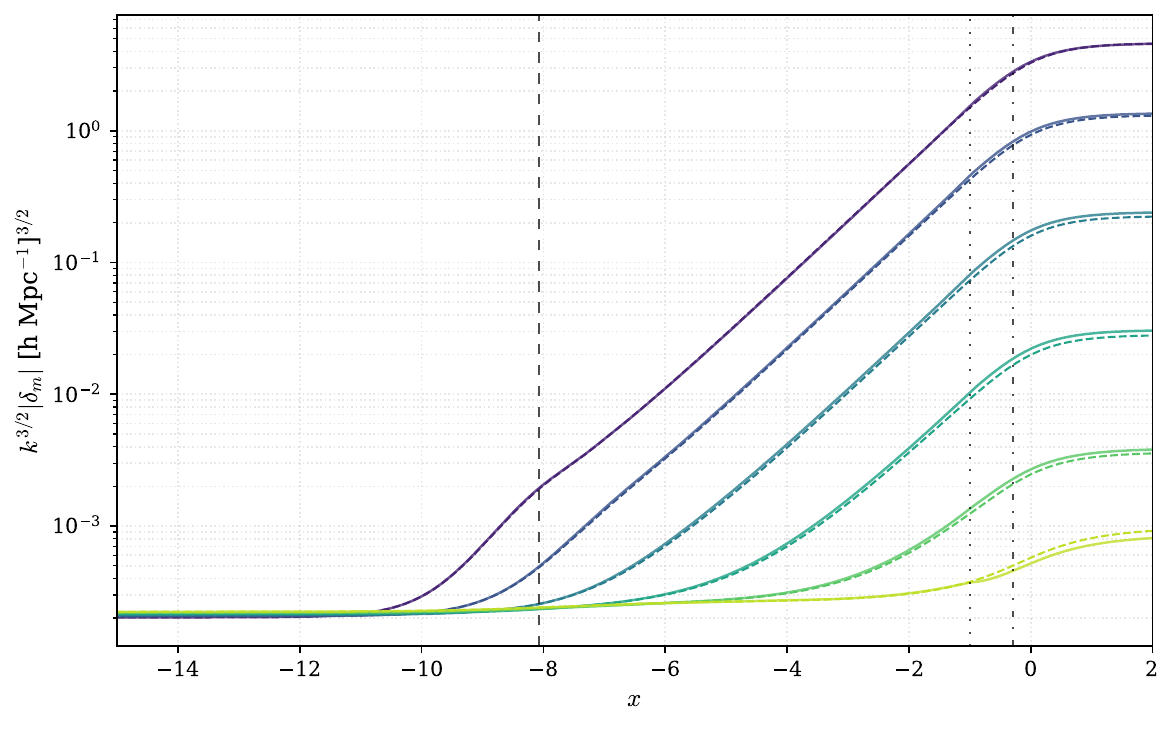} \\ \small (A) $\Lambda_s$CDM} &
\shortstack{\includegraphics[width=0.45\textwidth]{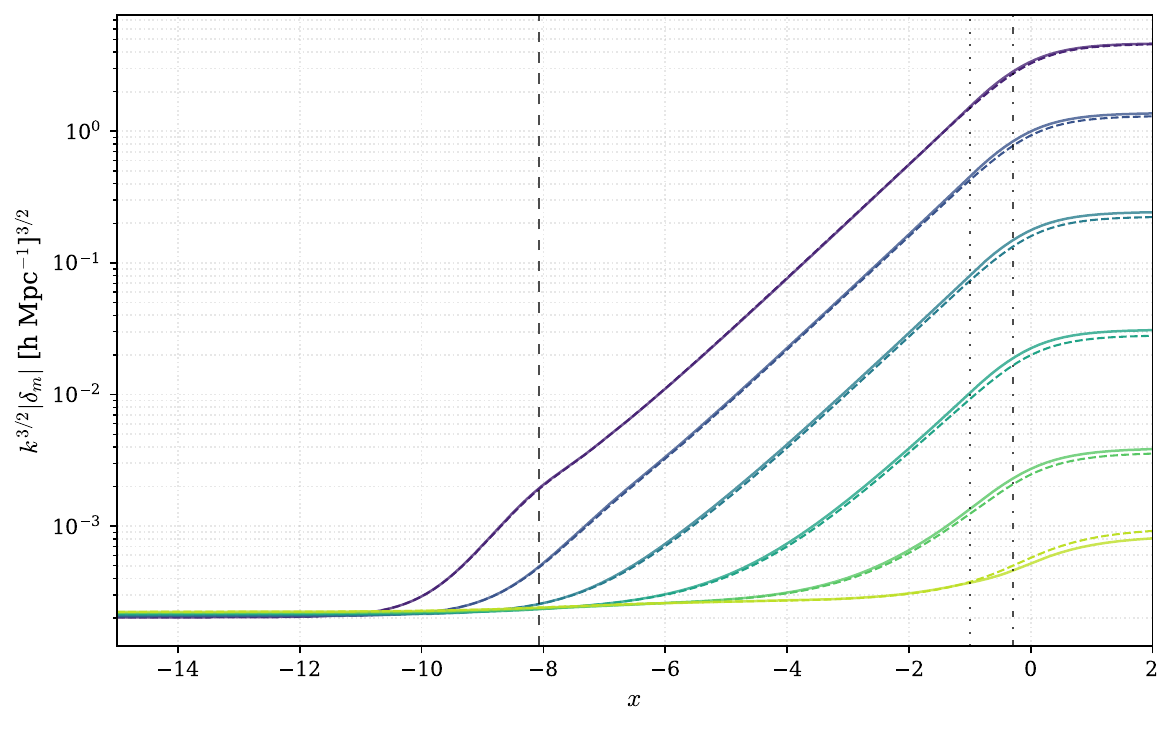} \\ \small (B) L$\Lambda$CDM} \\[0.5em]
\shortstack{\includegraphics[width=0.45\textwidth]{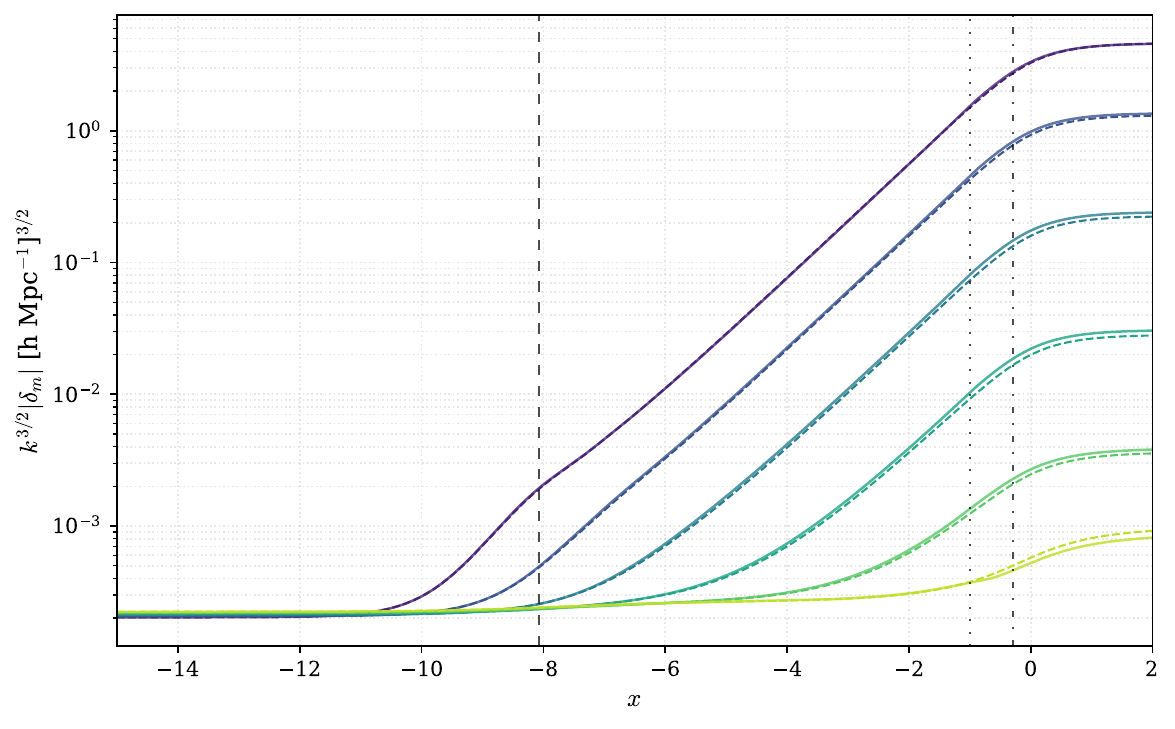} \\ \small (C) SSCDM} &
\shortstack{\includegraphics[width=0.45\textwidth]{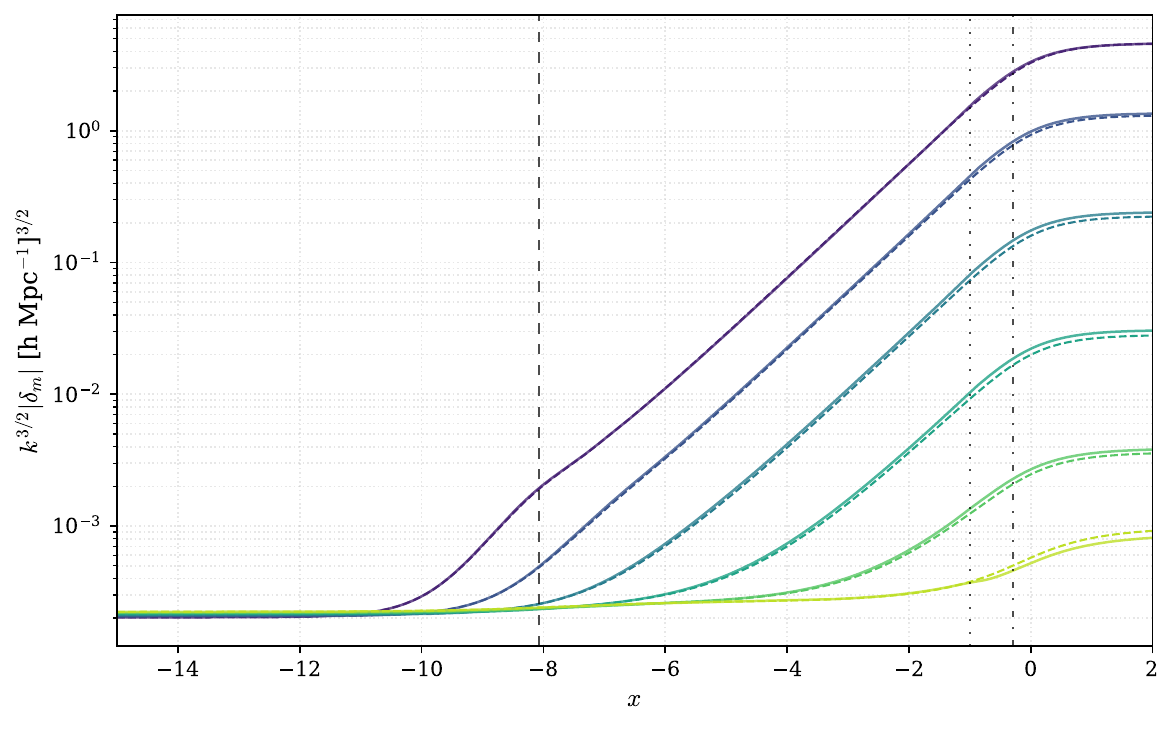} \\ \small (D) ECDM}
\end{tabular}

\caption{Evolution of fractional energy density perturbations in matter, $\delta_m$, for all sign-switching models (continuous lines) compared against $\Lambda$CDM (dashed lines). Each colour corresponds to a different Fourier mode:k = 3.33 $\times
10^{-4}$ h Mpc$^{-1}$ (Yellow), k = 1.04 $\times10^{-3}$ h Mpc$^{-1}$ (Green-yellow), k = 3.27$\times10^{-3}$ h Mpc$^{-1}$ (Green), k = $1.02 \times10^{-2}$ h Mpc$^{-1}$ (Cyan), k = 3.19$\times10^{-2}$ h Mpc$^{-1}$ (Blue), k = 0.1 h Mpc$^{-1}$ (Purple). The dashed left vertical line indicates the radiation-matter equality,  the right one the matter–DE equality, and the dotted vertical line between them corresponds to the sign-switching redshift.}
\label{fig:four-images-matter}
\end{figure}

In Fig.~\ref{fig:four-images-matter}, we present the evolution of cosmological perturbations from an initial epoch to a later stage in the Universe's evolution. The four panels depict the progression of the fractional energy density perturbations in matter, $\delta_m$, for the various models under investigation. In each panel, the solid lines denote the outcomes for the respective model, whereas the dotted lines indicate those of the $\Lambda$CDM model. Results are shown for six distinct wave numbers, with three qualitatively distinct behaviours depending on their range:

\begin{itemize}
    \item \textbf{Large $k$} = $0.1 $h Mpc$^{-1}$ (Purple) and k = $3.19 \times 10^{-2}$h Mpc$^{-1}$ (Blue).
    \item \textbf{Intermediate $k$} = $1.02 \times 10^{-2}$h Mpc$^{-1}$ (Cyan) and k = $3.27 \times 10^{-3}$h Mpc$^{-1}$ (Green).
    \item \textbf{Small $k$} = $1.04 \times 10^{-3}$h Mpc$^{-1}$  (Green-yellow) and k = $3.33 \times 10^{-4}$h Mpc$^{-1}$ (Yellow).
\end{itemize}
It is only the smaller modes that showcase a noticeable deviation from $\Lambda$CDM, that is, the modes that enter the horizon around the sign-switching. The behaviour is similar and consistent around all models. During the radiation-dominated era, the matter density contrast, $\delta_m$, remains approximately constant for each mode until it enters the horizon. Following horizon entry, gravitational collapse initiates the growth of perturbations, which becomes exponential during the matter-dominated epoch. In the subsequent DE-dominated phase, this growth ceases, and $\delta_m$ asymptotically approaches a constant value for each mode. It is also noteworthy that modes with larger wave numbers, $k$, enter the horizon earlier and, as a result, attain higher values of $\delta_m$ throughout cosmic evolution.

\subsubsection{Gravitational potential}
\begin{figure}[t]
\centering

\begin{tabular}{cccc}
\shortstack{\includegraphics[width=0.225\textwidth]{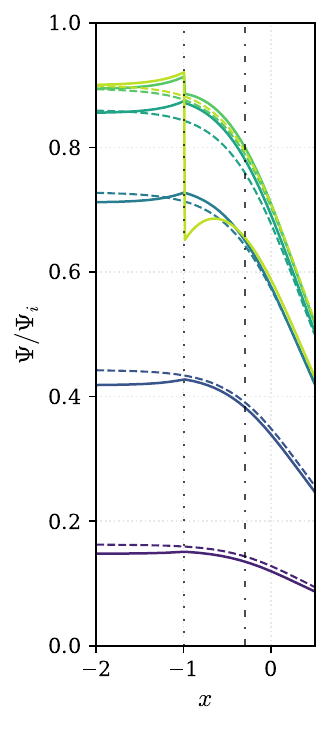} \\ \small (A) $\Lambda_s$CDM} &
\shortstack{\includegraphics[width=0.225\textwidth]{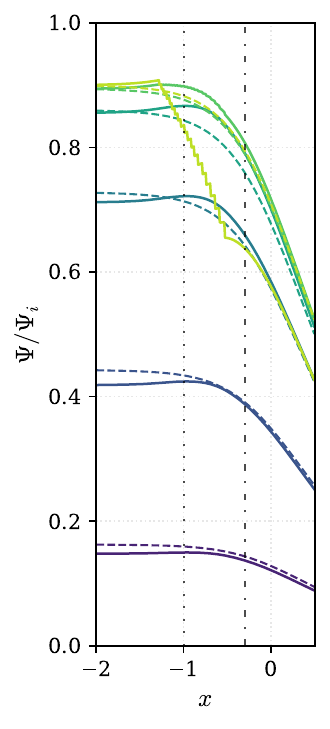} \\ \small (B) L$\Lambda$CDM} &
\shortstack{\includegraphics[width=0.225\textwidth]{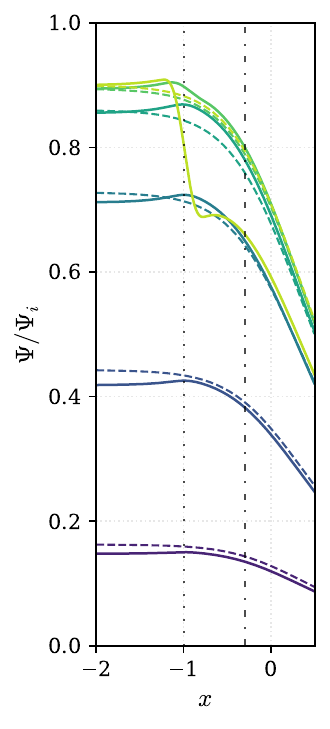} \\ \small (C) SSCDM} &
\shortstack{\includegraphics[width=0.225\textwidth]{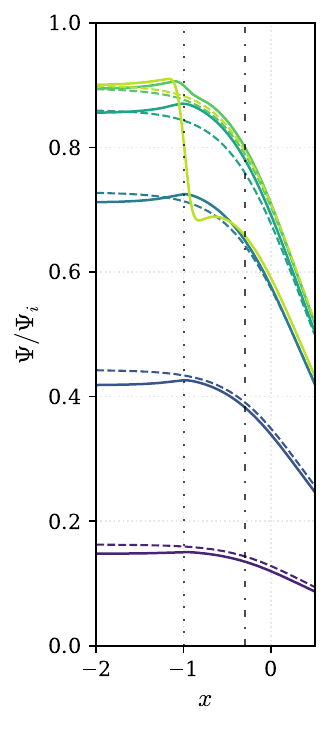} \\ \small (D) ECDM}
\end{tabular}

\caption{Gravitational potential $\Psi/\Psi_{\rm i}$, for all sign-switching models (continuous lines) compared against $\Lambda$CDM (dashed lines). Each colour corresponds to a different Fourier mode: k = 3.33 $\times
10^{-4}$ h Mpc$^{-1}$ (Yellow), k = 1.04 $\times10^{-3}$ h Mpc$^{-1}$ (Green-yellow), k = 3.27$\times10^{-3}$ h Mpc$^{-1}$ (Green), k = $1.02 \times10^{-2}$ h Mpc$^{-1}$ (Cyan), k = 3.19$\times10^{-2}$ h Mpc$^{-1}$ (Blue), k = 0.1 h Mpc$^{-1}$ (Purple). The dashed right vertical line indicates the matter–DE equality, and the dotted left vertical line corresponds to the sign-switching redshift.
}
\label{zoompsi}
\end{figure}

In Fig.~\ref{zoompsi}, we show the evolution of the gravitational potential, normalised to its initial value, $\Psi/\Psi_i$, for each model in comparison with $\Lambda$CDM. We consider the same wave modes as in the previous figures. For the sign-switching models, the evolution of the gravitational potential closely follows that of $\Lambda$CDM at all times, except in the vicinity of the sign transition. As the matter-dominated era draws to a close and the DE component becomes dynamically relevant, the negative DE density exerts a significant influence on the evolution, an effect that persists until the transition is completed and the DE density reaches its present value. In Ref.~\cite{Bouhmadi-Lopez:2025spo}, we investigated the evolution from the distant past ($z \sim 10^{6}$) to the far future ($z \sim -0.99$); however, in this proceedings contribution we only show the epoch during which the transition from negative to positive DE density occurs, as this is when the models exhibit the largest deviations from $\Lambda$CDM.

\begin{itemize}
    \item \textbf{Large $\boldsymbol{k}$}: The largest modes enter the horizon earliest, during the radiation-dominated era, and consequently experience the strongest suppression. They then settle into an almost constant configuration throughout the matter-dominated epoch. Around the epoch of the sign transition, a mild departure from the $\Lambda$CDM evolution becomes apparent, manifesting as a temporary enhancement sourced by the positive DE pressure prior to the transition. Once the sign change is completed, the potential undergoes a further period of decay over a timescale of a few e-folds, before converging towards a small residual constant value in  the future.

    \item \textbf{Intermediate $\boldsymbol{k}$}: The gravitational potential undergoes a brief decay around radiation–matter equality, after which it settles onto a plateau. Subsequently, in a manner analogous to the larger-scale modes, it exhibits an enhancement relative to $\Lambda$CDM as the sign-switching epoch is approached, followed by a phase of decay. Notably, the peak associated with the sign transition is more pronounced for these modes than for the larger ones.

    \item \textbf{Small $\boldsymbol{k}$}: As in the intermediate case, the modes experience a mild suppression around radiation–matter equality, after which a plateau is established. The most distinctive features arise as the sign-switching epoch is approached, where the behaviour becomes strongly model dependent. This is illustrated in Figs.~\ref{zoompsi}{\color{posurl}A} and~\ref{zoompsi}{\color{posurl}B}: model (A) undergoes a sharp drop at the transition, while model (B) exhibits a smoother evolution as a consequence of its step-like structure. In contrast, Models (C) and (D), shown in Fig.~\ref{zoompsi}, are characterised by a rapid but smooth decay associated with the continuous evolution of the DE density. At later times, the evolution converges towards the same values as the ones observed in the previous scenarios.

\end{itemize}

\subsubsection{\texorpdfstring{$f\sigma8$}{fs8} distribution \label{sec3d}}

One of the key quantities characterising structure formation is the growth rate of matter perturbations, defined as \cite{Albarran:2016mdu,Balcerzak:2012ae}
\begin{equation}
f = \frac{d \ln \delta_{\rm m}}{d \ln a},
\label{eq-f}
\end{equation}
where $\delta_{\rm m}$ denotes the fractional matter overdensity for a given Fourier mode. In observations, however, the combination $f\sigma_8$ is commonly used rather than $f$ alone. Here, $\sigma_8$ represents the root-mean-square of matter fluctuations within spheres of radius $8\,\text{h}^{-1}$Mpc and serves as the normalisation for the matter power spectrum. The advantage of considering $f\sigma_8$ lies in reducing the degeneracy between the linear galaxy bias $b$, which relates the dark matter perturbations to the observed galaxy density fluctuations, and $\sigma_8$, allowing for a more direct comparison with galaxy surveys.

\begin{figure}[H]
\centering
\begin{tabular}{c c}
% Row 1
\includegraphics[width=0.45\textwidth]{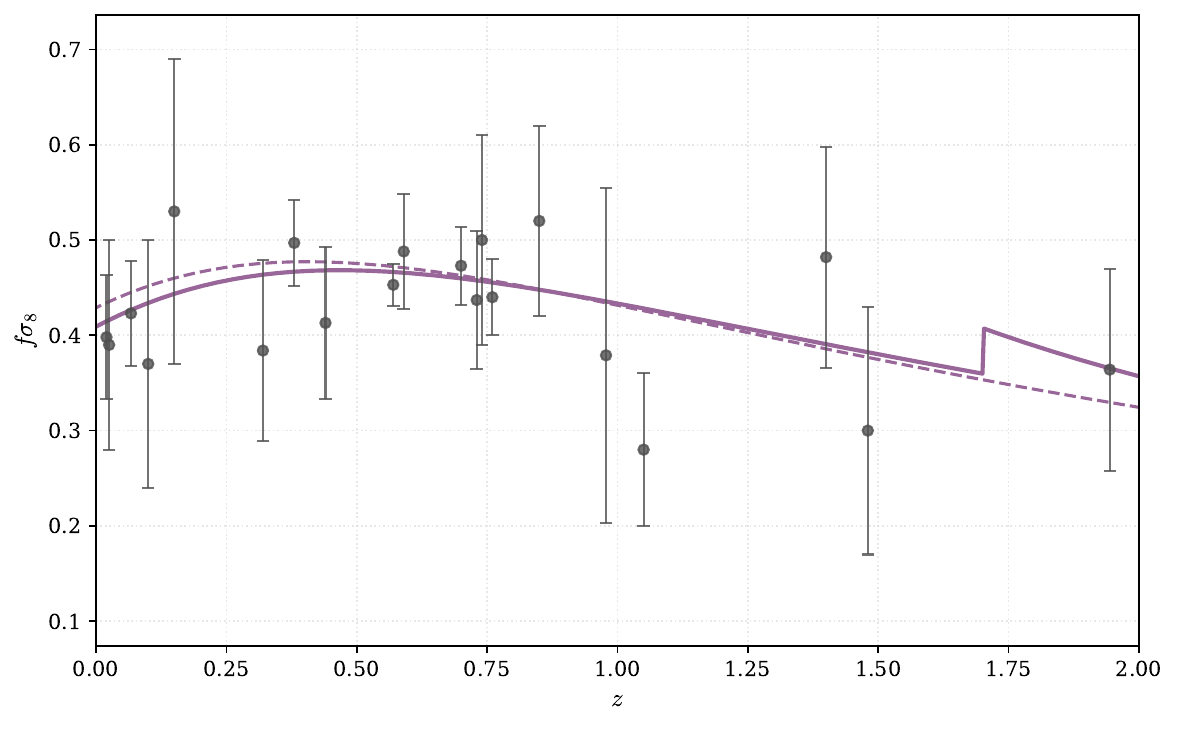}\hspace{-1em} & 
\includegraphics[width=0.45\textwidth]{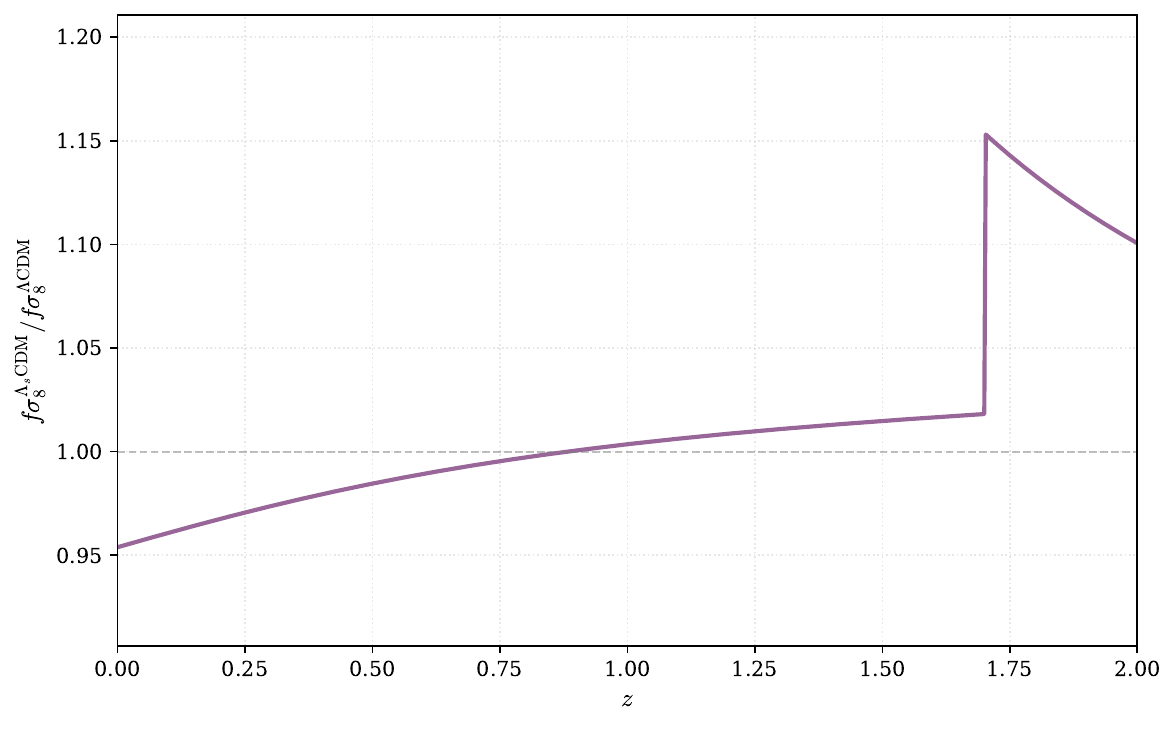} \\
\multicolumn{2}{c}{\small (A) $\Lambda_s$CDM} \\
[0.5em]
% Row 2  
\includegraphics[width=0.45\textwidth]{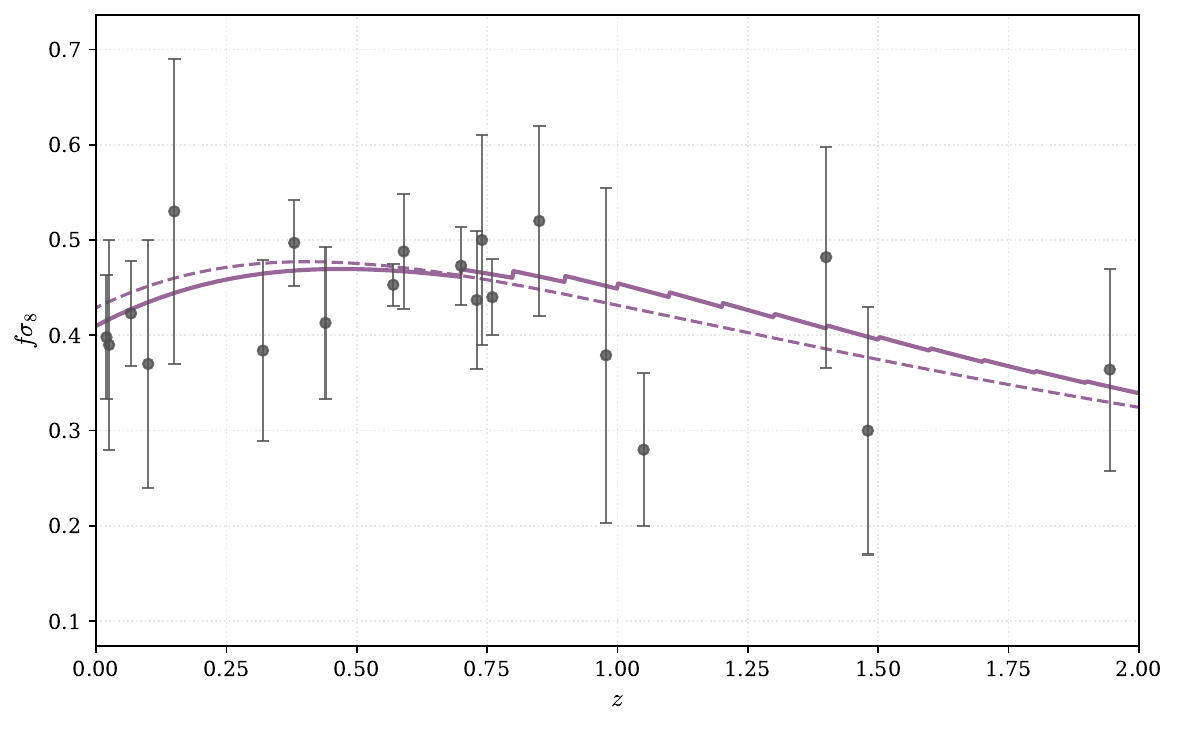}\hspace{-1em} & 
\includegraphics[width=0.45\textwidth]{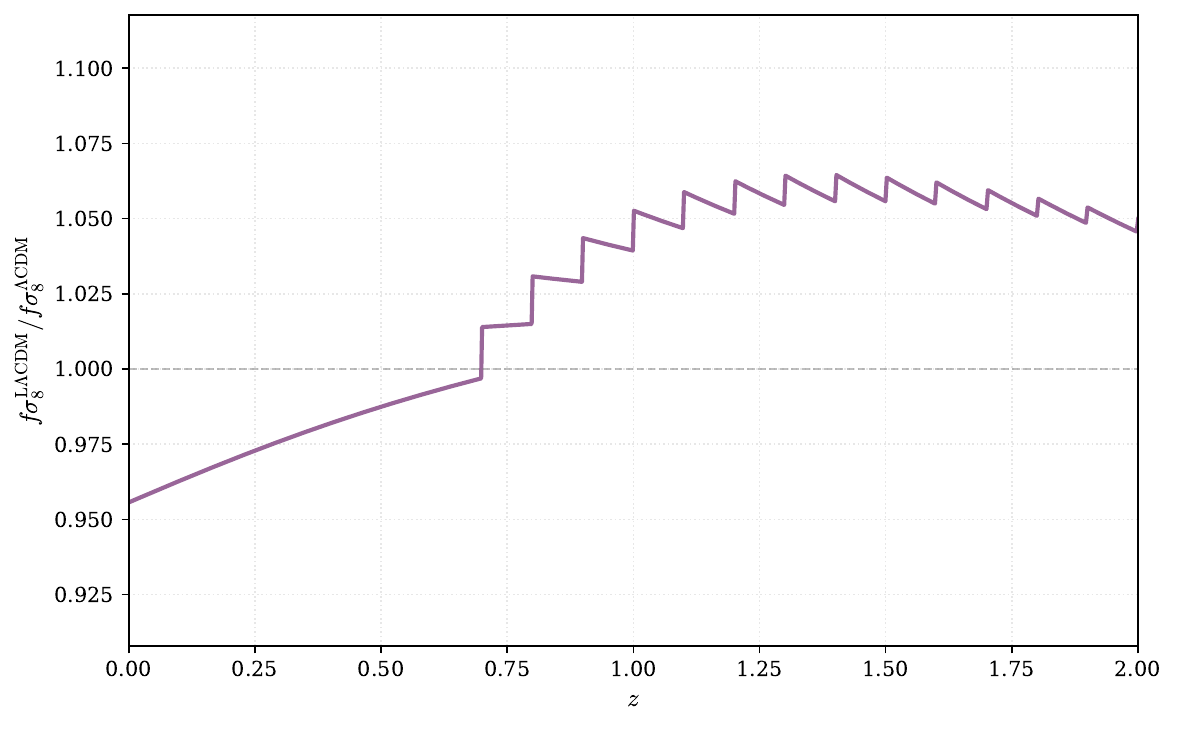} \\
\multicolumn{2}{c}{\small (B) L$\Lambda$CDM} \\
[0.5em]
% Row 3
\includegraphics[width=0.45\textwidth]{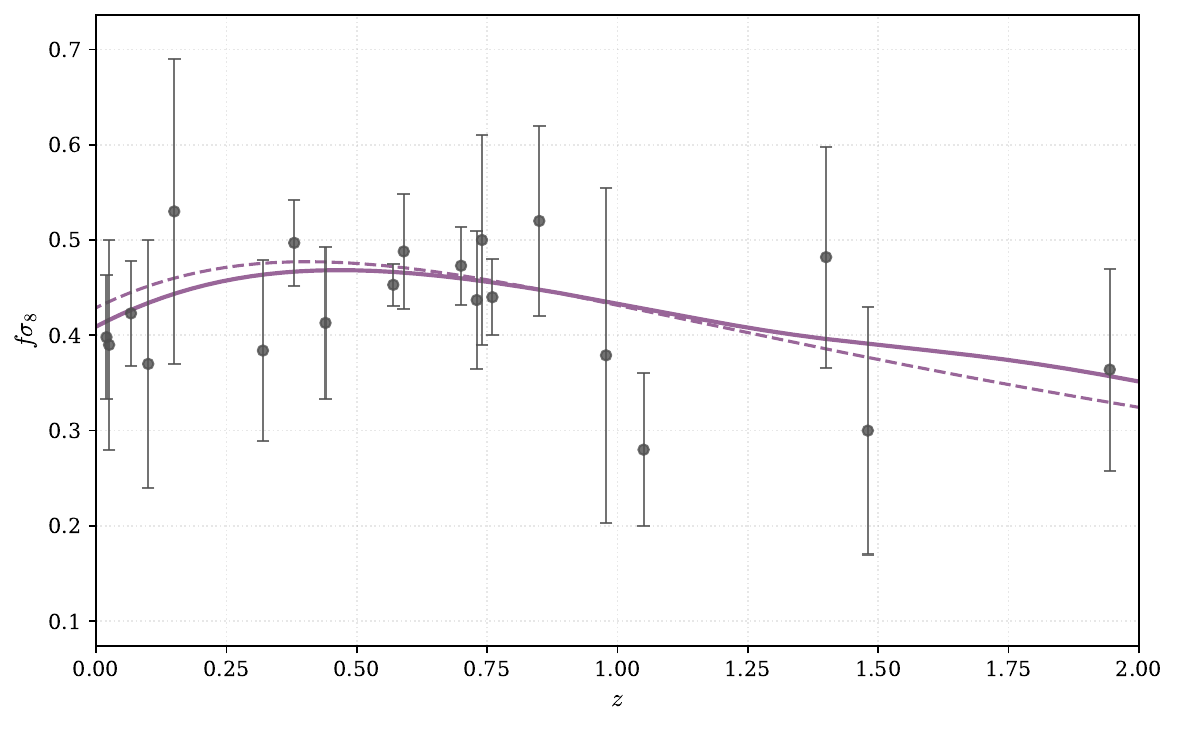}\hspace{-1em} & 
\includegraphics[width=0.45\textwidth]{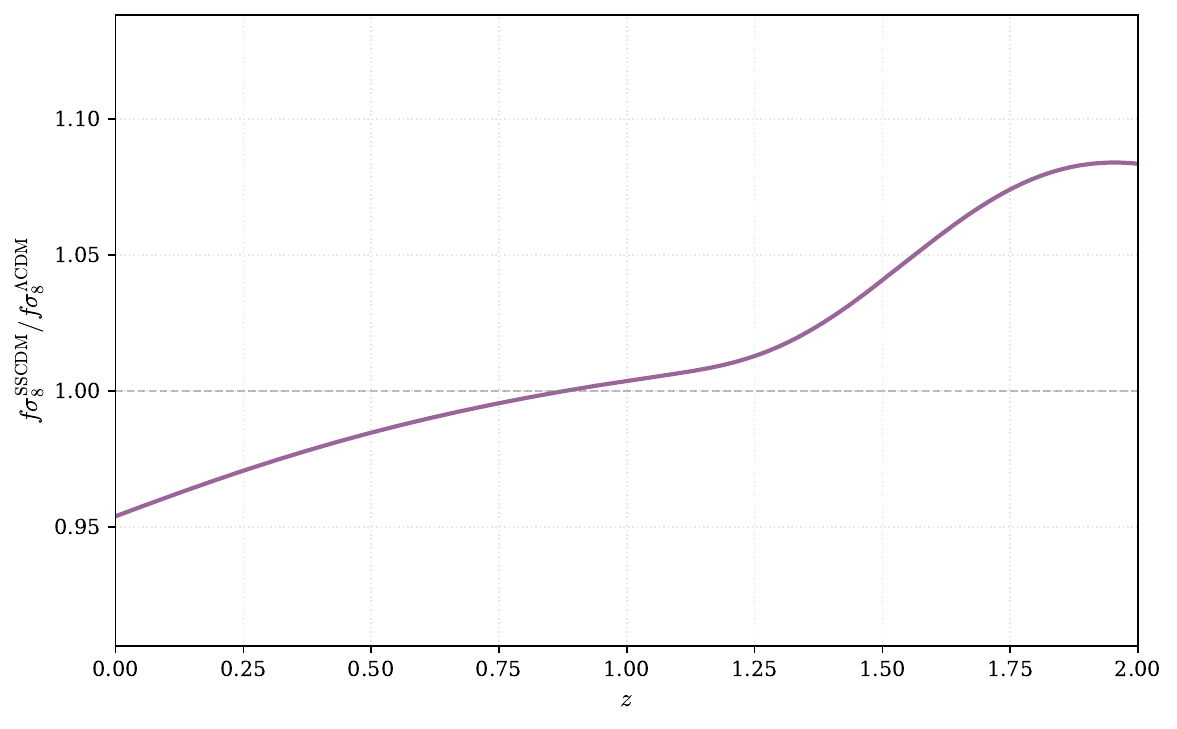} \\
\multicolumn{2}{c}{\small (C) SSCDM} \\
[0.5em]
% Row 4
\includegraphics[width=0.45\textwidth]{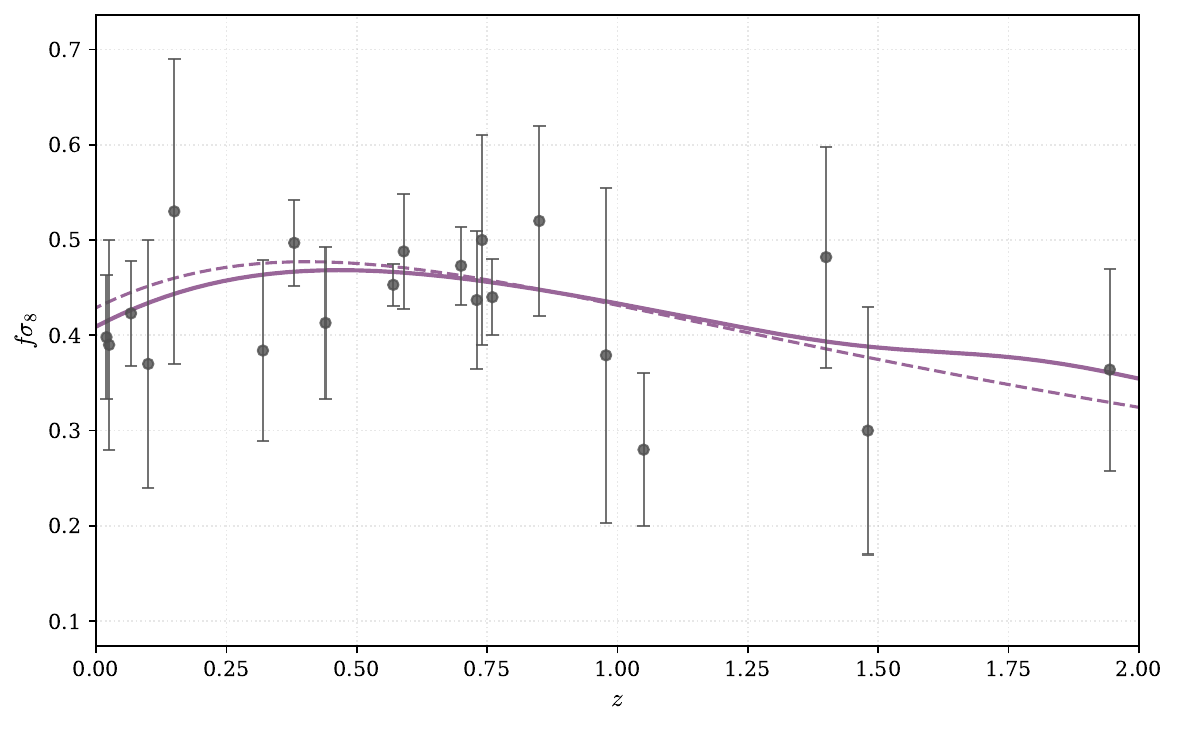}\hspace{-1em} & 
\includegraphics[width=0.45\textwidth]{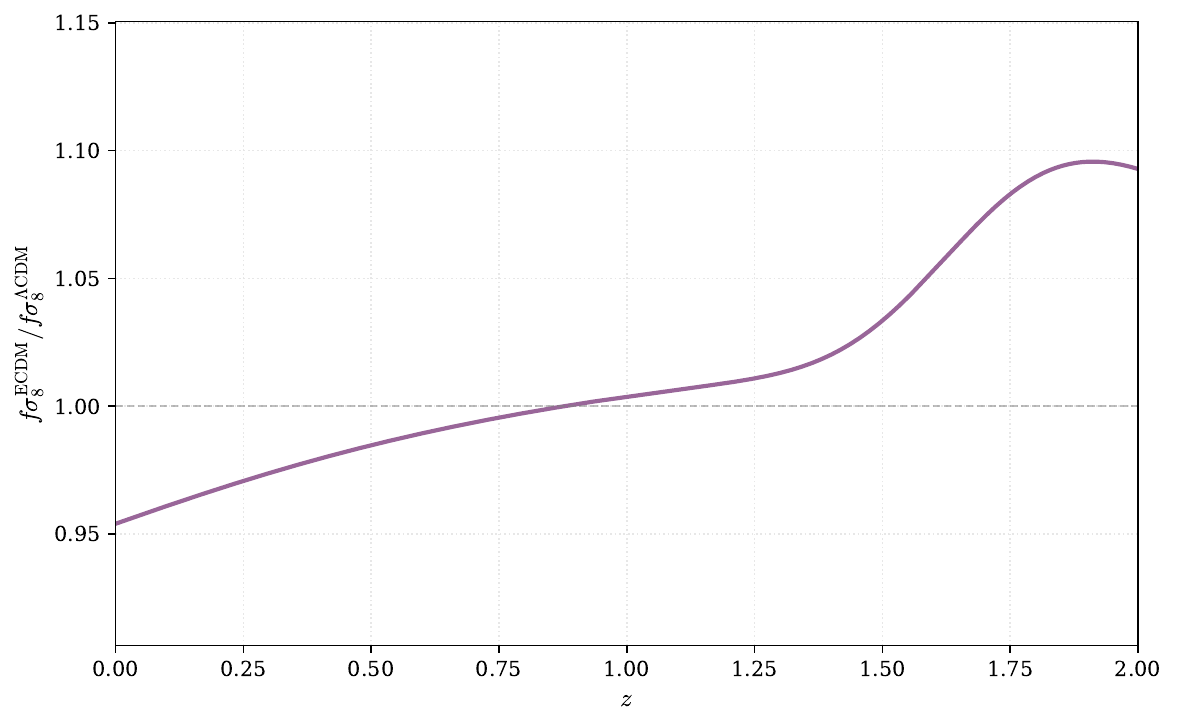} \\
\multicolumn{2}{c}{\small (D) ECDM}
\end{tabular}
\caption{The left columns corresponds to the evolution of $f \sigma_8$ for redshift between $z=0$ and $z=2$ against the data points of Ref.~\cite{Avila:2022xad}. The continuous lines correspond to the respective models of study while the dashed ones to $\Lambda$CDM. The right column corresponds to the ratio of $f\sigma8$ of each models against $\Lambda$CDM.}
\label{fig:all_plotsf8}
\end{figure}

The evolution of $\sigma_8$ is computed as
\begin{equation}
\sigma_8(z,k_{\sigma8}) = \sigma_8(0,k_{\sigma8}) \frac{\delta_{\rm m}(z,k_{\sigma8})}{\delta_{\rm m}(0,k_{\sigma8})},
\label{eq-sigma8}
\end{equation}
with $k_{\sigma_8} = 0.125\,\text{h\,Mpc}^{-1}$ corresponding to the physical $8\,\text{h}^{-1}$Mpc scale. We solve Eqs.~(\ref{perturb}) and (\ref{psieq}), together with (\ref{eq-f}) and (\ref{eq-sigma8}), to obtain the evolution of $f\sigma_8$ for each model. All computations adopt the Planck 2018 normalisation, $\sigma_8(0, k_{\sigma_8}) = 0.8120$ \cite{Planck:2018vyg}, and the results are compared both with the standard $\Lambda$CDM predictions and with observational data from Ref.~\cite{Avila:2022xad}.

Figure~\ref{fig:all_plotsf8} summarises these results, showing the evolution of $f\sigma_8$ relative to $\Lambda$CDM across $z\in[0,2]$ for all four models. Predictions of our models are displayed in continuous violet, while $\Lambda$CDM is shown in dashed lines. We find that all scenarios remain within the observational uncertainties throughout most of the redshift range. Notably, all four models provide an improved match to the highest and lowest redshift measurements from quasar and EBOSS data \cite{Avila:2022xad}, lying closer to the observed value than the $\Lambda$CDM curve.

Panels~\ref{fig:all_plotsf8}\textcolor{posurl}{A} and \ref{fig:all_plotsf8}\textcolor{posurl}{B} illustrate the difference between a single-step transition and a generalised ladder-step scenario, respectively. In the continuous-transition models (C) and (D) (panels \ref{fig:all_plotsf8}\textcolor{posurl}{C} and \ref{fig:all_plotsf8}\textcolor{posurl}{D}), $f\sigma_8$ evolves from slightly elevated values toward the $\Lambda$CDM expectation as the DE density transitions from negative to positive. In all four models, an enhancement of structure formation is apparent in the vicinity of the sign-switching epoch, reflecting the temporary influence of negative DE density on matter clustering. In general, the same behaviour is apparent for all four models: a suppression of $f\sigma_8$ at redshifts close to the present epoch, when the DE density becomes larger than in $\Lambda$CDM, and an enhancement at higher redshift values due to the effect of the positive DE pressure.

\section{Conclusions\label{sec5}}

Motivated by scenarios in which the cosmological constant is allowed to change sign,  we have investigated a set of DE models extending the $\Lambda_{\rm s}$CDM framework. These include a phenomenological ladder-step construction (Model (B), or L$\Lambda$CDM) and two dynamically evolving realisations, SSCDM and ECDM (Models (C) and (D)), in which the transition from negative to positive DE density is implemented smoothly. Together, these models provide a unified setting in which abrupt and continuous sign changes of the DE sector can be studied within a consistent cosmological framework.

At the level of the homogeneous background, we explored the expansion history using both analytical and numerical techniques, focusing on the evolution of the Hubble parameter and on cosmographic diagnostics. The models with discontinuous transitions display sharp features in the deceleration and higher-order kinematic parameters, whereas the smooth-transition scenarios are characterised by regular and continuous behaviour across the entire redshift range.

We then extended the analysis to linear perturbations in order to assess the implications for structure formation. Matter density perturbations remain close to their $\Lambda$CDM counterparts, with only small departures across all cases. More pronounced differences appear in the evolution of the gravitational potential around the epochs of matter--DE equality and sign transition. In particular, discontinuous models give rise to abrupt variations, while the continuous scenarios exhibit a rapid but smooth decay. This behaviour is reflected in late-time observables, where a modest enhancement of the growth-related quantity $f\sigma_8$ is found relative to $\Lambda$CDM, especially in the smoothly evolving models, where the negative DE phase temporarily amplifies clustering.

Taken together, our results show that sign-switching models yield a viable background expansion history while simultaneously inducing subtle, yet potentially testable, deviations in the growth of cosmic structures. Although the predicted effects are moderate, they lead to a distinctive phenomenology that motivates a direct confrontation with observations. A first dedicated statistical analysis of the models introduced in \cite{Bouhmadi-Lopez:2025ggl} and reviewed here, using current cosmological datasets, was performed in \cite{Ibarra-Uriondo:2026zbp}, where their observational viability was assessed. That study shows that sign-switching dark energy models constitute a consistent extension of $\Lambda$CDM, capable of alleviating existing cosmological tensions while producing characteristic dynamical signatures.

\section*{Acknowledgements}

M. B.-L. is supported by the Basque Foundation of Science Ikerbasque. Our work is supported by the Spanish Grant PID2023-149016NB-I00 funded by (MCIN/AEI/10.13039/501100011033 and by “ERDF A way of making Europe). The authors acknowledge the contribution of the COST Action CA21136 “Addressing observational tensions in cosmology with systematics and fundamental physics (CosmoVerse)”. This work is also supported by the Basque government Grant No. IT1628-22 (Spain) and in particular B. I. U. is funded through that Grant.

\bibliographystyle{JHEP}
\bibliography{bibliography}

@article{Bouhmadi-Lopez:2008ukq,
    author = "Bouhmadi-L\'opez, Mariam and Ferrera, Antonio",
    title = "{Crossing the cosmological constant line in a dilatonic brane-world model with and without curvature corrections}",
    eprint = "0807.4678",
    archivePrefix = "arXiv",
    primaryClass = "hep-th",
    doi = "10.1088/1475-7516/2008/10/011",
    journal = "JCAP",
    volume = "10",
    pages = "011",
    year = "2008"
}

@article{Morais:2016bev,
    author = "Morais, Jo{\~a}o and Bouhmadi-L{\'o}pez, Mariam and Sravan Kumar, K. and Marto, Jo{\~a}o and Tavakoli, Yaser",
    title = "{Interacting 3-form dark energy models: distinguishing interactions and avoiding the Little Sibling of the Big Rip}",
    eprint = "1608.01679",
    archivePrefix = "arXiv",
    primaryClass = "gr-qc",
    doi = "10.1016/j.dark.2016.11.002",
    journal = "Phys. Dark Univ.",
    volume = "15",
    pages = "7--30",
    year = "2017"
}

@article{BeltranJimenez:2019tme,
    author = "Beltr\'an Jim\'enez, Jose and Heisenberg, Lavinia and Koivisto, Tomi Sebastian and Pekar, Simon",
    title = "{Cosmology in $f(Q)$ geometry}",
    eprint = "1906.10027",
    archivePrefix = "arXiv",
    primaryClass = "gr-qc",
    doi = "10.1103/PhysRevD.101.103507",
    journal = "Phys. Rev. D",
    volume = "101",
    number = "10",
    pages = "103507",
    year = "2020"
}

@article{BeltranJimenez:2018vdo,
    author = "Beltr\'an Jim\'enez, Jose and Heisenberg, Lavinia and Koivisto, Tomi S.",
    title = "{Teleparallel Palatini theories}",
    eprint = "1803.10185",
    archivePrefix = "arXiv",
    primaryClass = "gr-qc",
    reportNumber = "NORDITA-2018-023, IFT-UAM/CSIC-18-035, IFT-UAM-CSIC-18-035",
    doi = "10.1088/1475-7516/2018/08/039",
    journal = "JCAP",
    volume = "08",
    pages = "039",
    year = "2018"
}

@article{Ayuso:2020dcu,
    author = "Ayuso, Ismael and Lazkoz, Ruth and Salzano, Vincenzo",
    title = "{Observational constraints on cosmological solutions of $f(Q)$ theories}",
    eprint = "2012.00046",
    archivePrefix = "arXiv",
    primaryClass = "astro-ph.CO",
    doi = "10.1103/PhysRevD.103.063505",
    journal = "Phys. Rev. D",
    volume = "103",
    number = "6",
    pages = "063505",
    year = "2021"
}

@misc{Ayuso:2025vkc,
    author = "Ayuso, Ismael and Bouhmadi-L\'opez, Mariam and Chen, Che-Yu and Chew, Xiao Yan and Dialektopoulos, Konstantinos and Ong, Yen Chin",
    title = "{Insights in $f(Q)$ cosmology: the relevance of the connection}",
    eprint = "2506.03506",
    archivePrefix = "arXiv",
    primaryClass = "gr-qc",
    reportNumber = "RIKEN-iTHEMS-Report-25",
    month = "6",
    year = "2025"
}

@article{Akarsu:2019hmw,
    author = {Akarsu, \"Ozg\"ur and Barrow, John D. and Escamilla, Luis A. and Vazquez, J. Alberto},
    title = "{Graduated dark energy: Observational hints of a spontaneous sign switch in the cosmological constant}",
    eprint = "1912.08751",
    archivePrefix = "arXiv",
    primaryClass = "astro-ph.CO",
    doi = "10.1103/PhysRevD.101.063528",
    journal = "Phys. Rev. D",
    volume = "101",
    number = "6",
    pages = "063528",
    year = "2020"
}

@misc{CosmoVerse:2025txj,
    author = "Di Valentino, Eleonora and others",
    collaboration = "CosmoVerse",
    title = "{The CosmoVerse White Paper: Addressing observational tensions in cosmology with systematics and fundamental physics}",
    eprint = "2504.01669",
    archivePrefix = "arXiv",
    primaryClass = "astro-ph.CO",
    month = "4",
    year = "2025"
}

@article{BorislavovVasilev:2022gpp,
    author = "Borislavov Vasilev, Teodor and Bouhmadi-L{\'o}pez, Mariam and Mart{\'\i}n-Moruno, Prado",
    title = "{Phantom attractors in kinetic gravity braiding theories: a dynamical system approach}",
    eprint = "2212.02547",
    archivePrefix = "arXiv",
    primaryClass = "gr-qc",
    doi = "10.1088/1475-7516/2023/06/026",
    journal = "JCAP",
    volume = "06",
    pages = "026",
    year = "2023"
}

@misc{Bouhmadi-Lopez:2026dte,
    author = "Bouhmadi-L{\'o}pez, Mariam and Boiza, Carlos G. and Petronikolou, Maria and Saridakis, Emmanuel N.",
    title = "{Modified Teleparallel $f(T)$ Gravity, DESI BAO and the $H_0$ Tension}",
    eprint = "2601.22225",
    archivePrefix = "arXiv",
    primaryClass = "gr-qc",
    month = "1",
    year = "2026"
}

@article{Nojiri:2005sx,
    author = "Nojiri, Shin'ichi and Odintsov, Sergei D. and Tsujikawa, Shinji",
    title = "{Properties of singularities in (phantom) dark energy universe}",
    eprint = "hep-th/0501025",
    archivePrefix = "arXiv",
    doi = "10.1103/PhysRevD.71.063004",
    journal = "Phys. Rev. D",
    volume = "71",
    pages = "063004",
    year = "2005"
}

@article{BorislavovVasilev:2024loq,
    author = "Borislavov Vasilev, Teodor and Bouhmadi-L\'opez, Mariam and Mart\'\i{}n-Moruno, Prado",
    title = "{Dark energy with a shift-symmetric scalar field: Obstacles, loophole hunting and dead ends}",
    eprint = "2406.12576",
    archivePrefix = "arXiv",
    primaryClass = "gr-qc",
    reportNumber = "IPARCOS-UCM-24-030",
    doi = "10.1016/j.dark.2024.101679",
    journal = "Phys. Dark Univ.",
    volume = "46",
    pages = "101679",
    year = "2024"
}

@article{Chiang:2025qxg,
    author = "Chiang, Hsu-Wen and Boiza, Carlos G. and Bouhmadi-L{\'o}pez, Mariam",
    title = "{Observational constraints on generalised axion-like potentials for the late Universe}",
    eprint = "2503.04898",
    archivePrefix = "arXiv",
    primaryClass = "astro-ph.CO",
    doi = "10.1088/1475-7516/2025/08/064",
    journal = "JCAP",
    volume = "08",
    pages = "064",
    year = "2025"
}

@article{Boiza:2025xpn,
    author = "Boiza, Carlos G. and Petronikolou, Maria and Bouhmadi-L{\'o}pez, Mariam and Saridakis, Emmanuel N.",
    title = "{Addressing H $_{0}$ and S $_{8}$ tensions within f(Q) cosmology}",
    eprint = "2505.18264",
    archivePrefix = "arXiv",
    primaryClass = "astro-ph.CO",
    doi = "10.1088/1475-7516/2025/12/011",
    journal = "JCAP",
    volume = "12",
    pages = "011",
    year = "2025"
}

@inproceedings{Brandenberger:1993zc,
    author = "Brandenberger, Robert H. and Feldman, H. and Mukhanov, Viatcheslav F.",
    title = "{Classical and quantum theory of perturbations in inflationary universe models}",
    booktitle = "{37th Yamada Conference: Evolution of the Universe and its Observational Quest}",
    eprint = "astro-ph/9307016",
    archivePrefix = "arXiv",
    reportNumber = "BROWN-HET-914",
    pages = "19--30",
    month = "7",
    year = "1993"
}

@book{2004,
   title={The Early Universe and Observational Cosmology},
   ISBN={9783540409182},
   ISSN={1616-6361},
   url={http://dx.doi.org/10.1007/b97189},
   journal={Lecture Notes in Physics},
   publisher={Springer Berlin Heidelberg},
   author={R.H.Brandenberger},
   year={2004} }

@article{Malik_2009,
   title={Cosmological perturbations},
   volume={475},
   ISSN={0370-1573},
   DOI={10.1016/j.physrep.2009.03.001},
   number={1–4},
   journal={Physics Reports},
   publisher={Elsevier BV},
   author={Malik, Karim A. and Wands, David},
   year={2009},
   month=may, pages={1–51} }

@book{bauman,
title={Cosmology: Part III Mathematical Tripos
(Lecture notes)},
url={https://www.damtp.cam.ac.uk/user/examples/3R2La.pdf},
publisher={Cambridge},
author={D. Baumann},
year={2012}
}

@article{Albarran:2016mdu,
    author = "Albarran, Imanol and Bouhmadi-L{\'o}pez, Mariam and Morais, Jo{\~a}o",
    title = "{Cosmological perturbations in an effective and genuinely phantom dark energy Universe}",
    eprint = "1611.00392",
    archivePrefix = "arXiv",
    primaryClass = "astro-ph.CO",
    doi = "10.1016/j.dark.2017.04.002",
    journal = "Phys. Dark Univ.",
    volume = "16",
    pages = "94--108",
    year = "2017"
}

@article{Ma:1995ey,
    author = "Ma, Chung-Pei and Bertschinger, Edmund",
    title = "{Cosmological perturbation theory in the synchronous and conformal Newtonian gauges}",
    eprint = "astro-ph/9506072",
    archivePrefix = "arXiv",
    doi = "10.1086/176550",
    journal = "Astrophys. J.",
    volume = "455",
    pages = "7--25",
    year = "1995"
}

@article{Planck:2018vyg,
    author = "Aghanim, N. and others",
    collaboration = "Planck",
    title = "{Planck 2018 results. VI. Cosmological parameters}",
    eprint = "1807.06209",
    archivePrefix = "arXiv",
    primaryClass = "astro-ph.CO",
    doi = "10.1051/0004-6361/201833910",
    journal = "Astron. Astrophys.",
    volume = "641",
    pages = "A6",
    year = "2020",
    note = "[Erratum: Astron.Astrophys. 652, C4 (2021)]"
}

@misc{Ghafari:2025eql,
    author = {Ghafari, Payam and Najafi, Mahdi and Ghodsi Yengejeh, Mina and {\"O}z{\"u}lker, Emre and Di Valentino, Eleonora and Firouzjaee, Javad T.},
    title = "{A Multi-Probe ISW Study of Dark Energy Models with Negative Energy Density: Galaxy Correlations, Lensing Bispectrum, and Planck ISW-Lensing Likelihood}",
    eprint = "2512.07060",
    archivePrefix = "arXiv",
    primaryClass = "astro-ph.CO",
    month = "12",
    year = "2025"
}

@article{Paraskevas:2024ytz,
    author = "Paraskevas, Evangelos A. and Cam, Arman and Perivolaropoulos, Leandros and Akarsu, Ozgur",
    title = "{Transition dynamics in the \ensuremath{\Lambda}sCDM model: Implications for bound cosmic structures}",
    eprint = "2402.05908",
    archivePrefix = "arXiv",
    primaryClass = "astro-ph.CO",
    doi = "10.1103/PhysRevD.109.103522",
    journal = "Phys. Rev. D",
    volume = "109",
    number = "10",
    pages = "103522",
    year = "2024"
}

@article{SupernovaSearchTeam:1998fmf,
    author = "Riess, Adam G. and others",
    collaboration = "Supernova Search Team",
    title = "{Observational evidence from supernovae for an accelerating universe and a cosmological constant}",
    eprint = "astro-ph/9805201",
    archivePrefix = "arXiv",
    doi = "10.1086/300499",
    journal = "Astron. J.",
    volume = "116",
    pages = "1009--1038",
    year = "1998"
}

@article{DiValentino:2021izs,
    author = "Di Valentino, Eleonora and Mena, Olga and Pan, Supriya and Visinelli, Luca and Yang, Weiqiang and Melchiorri, Alessandro and Mota, David F. and Riess, Adam G. and Silk, Joseph",
    title = "{In the realm of the Hubble tension\textemdash{}a review of solutions}",
    eprint = "2103.01183",
    archivePrefix = "arXiv",
    primaryClass = "astro-ph.CO",
    reportNumber = "IPPP/20/108",
    doi = "10.1088/1361-6382/ac086d",
    journal = "Class. Quant. Grav.",
    volume = "38",
    number = "15",
    pages = "153001",
    year = "2021"
}

@article{Poulin:2018,
    author = "Poulin, Vivian and Smith, Tristan L. and Karwal, Tanvi and Kamionkowski, Marc",
    title = "{Early Dark Energy Can Resolve The Hubble Tension}",
    eprint = "1811.04083",
    archivePrefix = "arXiv",
    primaryClass = "astro-ph.CO",
    doi = "10.1103/PhysRevLett.122.221301",
    journal = "Phys. Rev. Lett.",
    volume = "122",
    number = "22",
    pages = "221301",
    year = "2019"
}

@article{Kamionkowski:2022pkx,
    author = "Kamionkowski, Marc and Riess, Adam G.",
    title = "{The Hubble Tension and Early Dark Energy}",
    eprint = "2211.04492",
    archivePrefix = "arXiv",
    primaryClass = "astro-ph.CO",
    doi = "10.1146/annurev-nucl-111422-024107",
    journal = "Ann. Rev. Nucl. Part. Sci.",
    volume = "73",
    pages = "153--180",
    year = "2023"
}

@article{Perivolaropoulos:2021jda,
    author = "Perivolaropoulos, Leandros and Skara, Foteini",
    title = "{Challenges for \ensuremath{\Lambda}CDM: An update}",
    eprint = "2105.05208",
    archivePrefix = "arXiv",
    primaryClass = "astro-ph.CO",
    doi = "10.1016/j.newar.2022.101659",
    journal = "New Astron. Rev.",
    volume = "95",
    pages = "101659",
    year = "2022"
}

@article{Abdalla:2022yfr,
    author = "Abdalla, Elcio and others",
    title = "{Cosmology intertwined: A review of the particle physics, astrophysics, and cosmology associated with the cosmological tensions and anomalies}",
    eprint = "2203.06142",
    archivePrefix = "arXiv",
    primaryClass = "astro-ph.CO",
    reportNumber = "FERMILAB-CONF-22-192-SCD",
    doi = "10.1016/j.jheap.2022.04.002",
    journal = "JHEAp",
    volume = "34",
    pages = "49--211",
    year = "2022"
}

@article{Kobayashi:2019hrl,
    author = "Kobayashi, Tsutomu",
    title = "{Horndeski theory and beyond: a review}",
    eprint = "1901.07183",
    archivePrefix = "arXiv",
    primaryClass = "gr-qc",
    reportNumber = "RUP-19-3",
    doi = "10.1088/1361-6633/ab2429",
    journal = "Rept. Prog. Phys.",
    volume = "82",
    number = "8",
    pages = "086901",
    year = "2019"
}

@article{Sotiriou:2008rp,
    author = "Sotiriou, Thomas P. and Faraoni, Valerio",
    title = "{f(R) Theories Of Gravity}",
    eprint = "0805.1726",
    archivePrefix = "arXiv",
    primaryClass = "gr-qc",
    doi = "10.1103/RevModPhys.82.451",
    journal = "Rev. Mod. Phys.",
    volume = "82",
    pages = "451--497",
    year = "2010"
}

@article{Ferraro:2006jd,
    author = "Ferraro, Rafael and Fiorini, Franco",
    title = "{Modified teleparallel gravity: Inflation without inflaton}",
    eprint = "gr-qc/0610067",
    archivePrefix = "arXiv",
    doi = "10.1103/PhysRevD.75.084031",
    journal = "Phys. Rev. D",
    volume = "75",
    pages = "084031",
    year = "2007"
}

@article{Maldacena:1997re,
    author = "Maldacena, Juan Martin",
    title = "{The Large N limit of superconformal field theories and supergravity}",
    eprint = "hep-th/9711200",
    archivePrefix = "arXiv",
    reportNumber = "HUTP-97-A097, HUTP-98-A097",
    doi = "10.4310/ATMP.1998.v2.n2.a1",
    journal = "Adv. Theor. Math. Phys.",
    volume = "2",
    pages = "231--252",
    year = "1998"
}

@article{Avila:2022xad,
    author = "Avila, Felipe and Bernui, Armando and Bonilla, Alexander and Nunes, Rafael C.",
    title = "{Inferring $S_8(z)$ and $\gamma (z)$ with cosmic growth rate measurements using machine learning}",
    eprint = "2201.07829",
    archivePrefix = "arXiv",
    primaryClass = "astro-ph.CO",
    doi = "10.1140/epjc/s10052-022-10561-0",
    journal = "Eur. Phys. J. C",
    volume = "82",
    number = "7",
    pages = "594",
    year = "2022"
}

@article{Visser:2004bf,
    author = "Visser, Matt",
    editor = "McClelland, D. E. and Scott, S. M.",
    title = "{Cosmography: Cosmology without the Einstein equations}",
    eprint = "gr-qc/0411131",
    archivePrefix = "arXiv",
    doi = "10.1007/s10714-005-0134-8",
    journal = "Gen. Rel. Grav.",
    volume = "37",
    pages = "1541--1548",
    year = "2005"
}

@article{Cattoen:2007sk,
    author = "Cattoen, Celine and Visser, Matt",
    title = "{The Hubble series: Convergence properties and redshift variables}",
    eprint = "0710.1887",
    archivePrefix = "arXiv",
    primaryClass = "gr-qc",
    doi = "10.1088/0264-9381/24/23/018",
    journal = "Class. Quant. Grav.",
    volume = "24",
    pages = "5985--5998",
    year = "2007"
}

@article{Capozziello:2008qc,
    author = "Capozziello, S. and Cardone, V. F. and Salzano, V.",
    title = "{Cosmography of f(R) gravity}",
    eprint = "0802.1583",
    archivePrefix = "arXiv",
    primaryClass = "astro-ph",
    doi = "10.1103/PhysRevD.78.063504",
    journal = "Phys. Rev. D",
    volume = "78",
    pages = "063504",
    year = "2008"
}

@article{Favale:2023lnp,
    author = "Favale, Arianna and G\'omez-Valent, Adri\`a and Migliaccio, Marina",
    title = "{Cosmic chronometers to calibrate the ladders and measure the curvature of the Universe. A model-independent study}",
    eprint = "2301.09591",
    archivePrefix = "arXiv",
    primaryClass = "astro-ph.CO",
    doi = "10.1093/mnras/stad1621",
    journal = "Mon. Not. Roy. Astron. Soc.",
    volume = "523",
    number = "3",
    pages = "3406--3422",
    year = "2023"
}

@article{Harko:2011kv,
    author = "Harko, Tiberiu and Lobo, Francisco S. N. and Nojiri, Shin'ichi and Odintsov, Sergei D.",
    title = "{$f(R,T)$ gravity}",
    eprint = "1104.2669",
    archivePrefix = "arXiv",
    primaryClass = "gr-qc",
    doi = "10.1103/PhysRevD.84.024020",
    journal = "Phys. Rev. D",
    volume = "84",
    pages = "024020",
    year = "2011"
}

@article{Akarsu:2024qsi,
    author = {Akarsu, {\"O}zg{\"u}r and De Felice, Antonio and Di Valentino, Eleonora and Kumar, Suresh and Nunes, Rafael C. and {\"O}z{\"u}lker, Emre and Vazquez, J. Alberto and Yadav, Anita},
    title = "{{\ensuremath{\Lambda}}sCDM cosmology from a type-II minimally modified gravity}",
    eprint = "2402.07716",
    archivePrefix = "arXiv",
    primaryClass = "astro-ph.CO",
    reportNumber = "YITP-24-18",
    doi = "10.1093/mnras/staf2276",
    journal = "Mon. Not. Roy. Astron. Soc.",
    volume = "546",
    number = "1",
    pages = "staf2276",
    year = "2026"
}

@article{Niedermann:2019olb,
    author = "Niedermann, Florian and Sloth, Martin S.",
    title = "{New early dark energy}",
    eprint = "1910.10739",
    archivePrefix = "arXiv",
    primaryClass = "astro-ph.CO",
    doi = "10.1103/PhysRevD.103.L041303",
    journal = "Phys. Rev. D",
    volume = "103",
    number = "4",
    pages = "L041303",
    year = "2021"
}

@article{Ye:2021iwa,
    author = "Ye, Gen and Zhang, Jun and Piao, Yun-Song",
    title = "{Alleviating both H0 and S8 tensions: Early dark energy lifts the CMB-lockdown on ultralight axion}",
    eprint = "2107.13391",
    archivePrefix = "arXiv",
    primaryClass = "astro-ph.CO",
    doi = "10.1016/j.physletb.2023.137770",
    journal = "Phys. Lett. B",
    volume = "839",
    pages = "137770",
    year = "2023"
}

@article{Cruz:2023lmn,
    author = "Cruz, Juan S. and Niedermann, Florian and Sloth, Martin S.",
    title = "{Cold New Early Dark Energy pulls the trigger on the H$_{0}$ and S $_{8}$ tensions: a simultaneous solution to both tensions without new ingredients}",
    eprint = "2305.08895",
    archivePrefix = "arXiv",
    primaryClass = "astro-ph.CO",
    doi = "10.1088/1475-7516/2023/11/033",
    journal = "JCAP",
    volume = "11",
    pages = "033",
    year = "2023"
}

@misc{Niedermann:2023ssr,
    author = "Niedermann, Florian and Sloth, Martin S.",
    title = "{New Early Dark Energy as a solution to the $H_0$ and $S_8$ tensions}",
    eprint = "2307.03481",
    archivePrefix = "arXiv",
    primaryClass = "hep-ph",
    month = "7",
    year = "2023"
}

@article{Ye:2020btb,
    author = "Ye, Gen and Piao, Yun-Song",
    title = "{Is the Hubble tension a hint of AdS phase around recombination?}",
    eprint = "2001.02451",
    archivePrefix = "arXiv",
    primaryClass = "astro-ph.CO",
    doi = "10.1103/PhysRevD.101.083507",
    journal = "Phys. Rev. D",
    volume = "101",
    number = "8",
    pages = "083507",
    year = "2020"
}

@article{Ye:2020oix,
    author = "Ye, Gen and Piao, Yun-Song",
    title = "{$T_0$ censorship of early dark energy and AdS vacua}",
    eprint = "2008.10832",
    archivePrefix = "arXiv",
    primaryClass = "astro-ph.CO",
    doi = "10.1103/PhysRevD.102.083523",
    journal = "Phys. Rev. D",
    volume = "102",
    number = "8",
    pages = "083523",
    year = "2020"
}

@misc{Akarsu:2025gwi,
    author = {Akarsu, \"Ozg\"ur and Perivolaropoulos, Leandros and Tsikoundoura, Anna and Y\"ukselci, A. Emrah and Zhuk, Alexander},
    title = "{Dynamical dark energy with AdS-to-dS and dS-to-dS transitions: Implications for the $H_0$ tension}",
    eprint = "2502.14667",
    archivePrefix = "arXiv",
    primaryClass = "astro-ph.CO",
    month = "2",
    year = "2025"
}

@article{DeFelice:2010aj,
    author = "De Felice, Antonio and Tsujikawa, Shinji",
    title = "{f(R) theories}",
    eprint = "1002.4928",
    archivePrefix = "arXiv",
    primaryClass = "gr-qc",
    doi = "10.12942/lrr-2010-3",
    journal = "Living Rev. Rel.",
    volume = "13",
    pages = "3",
    year = "2010"
}

@article{Nojiri:2004bi,
    author = "Nojiri, Shin'ichi and Odintsov, Sergei D.",
    title = "{Gravity assisted dark energy dominance and cosmic acceleration}",
    eprint = "astro-ph/0403622",
    archivePrefix = "arXiv",
    doi = "10.1016/j.physletb.2004.08.045",
    journal = "Phys. Lett. B",
    volume = "599",
    pages = "137--142",
    year = "2004"
}

@article{DiValentino:2020naf,
    author = "Di Valentino, Eleonora and Mukherjee, Ankan and Sen, Anjan A.",
    title = "{Dark Energy with Phantom Crossing and the $H_0$ Tension}",
    eprint = "2005.12587",
    archivePrefix = "arXiv",
    primaryClass = "astro-ph.CO",
    reportNumber = "IPPP/20/89",
    doi = "10.3390/e23040404",
    journal = "Entropy",
    volume = "23",
    number = "4",
    pages = "404",
    year = "2021"
}

@article{Alestas:2020mvb,
    author = "Alestas, G. and Kazantzidis, L. and Perivolaropoulos, L.",
    title = "{$H_0$ tension, phantom dark energy, and cosmological parameter degeneracies}",
    eprint = "2004.08363",
    archivePrefix = "arXiv",
    primaryClass = "astro-ph.CO",
    doi = "10.1103/PhysRevD.101.123516",
    journal = "Phys. Rev. D",
    volume = "101",
    number = "12",
    pages = "123516",
    year = "2020"
}

@article{Adil:2023exv,
    author = {Adil, Shahnawaz A. and Akarsu, \"Ozg\"ur and Di Valentino, Eleonora and Nunes, Rafael C. and \"Oz\"ulker, Emre and Sen, Anjan A. and Specogna, Enrico},
    title = "{Omnipotent dark energy: A phenomenological answer to the Hubble tension}",
    eprint = "2306.08046",
    archivePrefix = "arXiv",
    primaryClass = "astro-ph.CO",
    doi = "10.1103/PhysRevD.109.023527",
    journal = "Phys. Rev. D",
    volume = "109",
    number = "2",
    pages = "023527",
    year = "2024"
}

@article{Kumar:2017dnp,
    author = "Kumar, Suresh and Nunes, Rafael C.",
    title = "{Echo of interactions in the dark sector}",
    eprint = "1702.02143",
    archivePrefix = "arXiv",
    primaryClass = "astro-ph.CO",
    doi = "10.1103/PhysRevD.96.103511",
    journal = "Phys. Rev. D",
    volume = "96",
    number = "10",
    pages = "103511",
    year = "2017"
}

@article{Nunes:2022bhn,
    author = "Nunes, Rafael C. and Vagnozzi, Sunny and Kumar, Suresh and Di Valentino, Eleonora and Mena, Olga",
    title = "{New tests of dark sector interactions from the full-shape galaxy power spectrum}",
    eprint = "2203.08093",
    archivePrefix = "arXiv",
    primaryClass = "astro-ph.CO",
    doi = "10.1103/PhysRevD.105.123506",
    journal = "Phys. Rev. D",
    volume = "105",
    number = "12",
    pages = "123506",
    year = "2022"
}

@article{Akarsu:2024nas,
    author = "Akarsu, Ozgur and Bulduk, Bilal and De Felice, Antonio and Kat{\i}rc{\i}, Nihan and Uzun, N. Merve",
    title = "{Unexplored regions in teleparallel f(T) gravity: Sign-changing dark energy density}",
    eprint = "2410.23068",
    archivePrefix = "arXiv",
    primaryClass = "gr-qc",
    reportNumber = "YITP-24-119",
    doi = "10.1103/1xd4-k91h",
    journal = "Phys. Rev. D",
    volume = "112",
    number = "8",
    pages = "083532",
    year = "2025"
}

@article{Balcerzak:2012ae,
    author = "Balcerzak, Adam and Denkiewicz, Tomasz",
    title = "{Density preturbations in a finite scale factor singularity universe}",
    eprint = "1202.3280",
    archivePrefix = "arXiv",
    primaryClass = "astro-ph.CO",
    doi = "10.1103/PhysRevD.86.023522",
    journal = "Phys. Rev. D",
    volume = "86",
    pages = "023522",
    year = "2012"
}

@article{Akarsu:2021fol,
    author = {Akarsu, \"Ozg\"ur and Kumar, Suresh and \"Oz\"ulker, Emre and Vazquez, J. Alberto},
    title = "{Relaxing cosmological tensions with a sign switching cosmological constant}",
    eprint = "2108.09239",
    archivePrefix = "arXiv",
    primaryClass = "astro-ph.CO",
    doi = "10.1103/PhysRevD.104.123512",
    journal = "Phys. Rev. D",
    volume = "104",
    number = "12",
    pages = "123512",
    year = "2021"
}

@article{Akarsu:2022typ,
    author = {Akarsu, Ozgur and Kumar, Suresh and \"Oz\"ulker, Emre and Vazquez, J. Alberto and Yadav, Anita},
    title = "{Relaxing cosmological tensions with a sign switching cosmological constant: Improved results with Planck, BAO, and Pantheon data}",
    eprint = "2211.05742",
    archivePrefix = "arXiv",
    primaryClass = "astro-ph.CO",
    doi = "10.1103/PhysRevD.108.023513",
    journal = "Phys. Rev. D",
    volume = "108",
    number = "2",
    pages = "023513",
    year = "2023"
}

@misc{Akarsu:2023mfb,
    author = "Akarsu, Ozgur and Di Valentino, Eleonora and Kumar, Suresh and Nunes, Rafael C. and Vazquez, J. Alberto and Yadav, Anita",
    title = "{$\Lambda_{\rm s}$CDM model: A promising scenario for alleviation of cosmological tensions}",
    eprint = "2307.10899",
    archivePrefix = "arXiv",
    primaryClass = "astro-ph.CO",
    month = "7",
    year = "2023"
}

@article{Akarsu:2024eoo,
    author = {Akarsu, \"Ozg\"ur and De Felice, Antonio and Di Valentino, Eleonora and Kumar, Suresh and Nunes, Rafael C. and \"Oz\"ulker, Emre and Vazquez, J. Alberto and Yadav, Anita},
    title = "{Cosmological constraints on \ensuremath{\Lambda}sCDM scenario in a type II minimally modified gravity}",
    eprint = "2406.07526",
    archivePrefix = "arXiv",
    primaryClass = "astro-ph.CO",
    reportNumber = "YITP-24-57",
    doi = "10.1103/PhysRevD.110.103527",
    journal = "Phys. Rev. D",
    volume = "110",
    number = "10",
    pages = "103527",
    year = "2024"
}

@article{Yadav:2025vpx,
    author = "Yadav, Manish and Dixit, Archana and Pradhan, Anirudh and Barak, M. S.",
    title = "{Empirical validation: Investigating the {\ensuremath{\Lambda}}sCDM model with new DESI BAO observations}",
    eprint = "2509.26049",
    archivePrefix = "arXiv",
    primaryClass = "astro-ph.CO",
    doi = "10.1016/j.jheap.2025.100453",
    journal = "JHEAp",
    volume = "49",
    pages = "100453",
    year = "2026"
}

@article{Souza:2024qwd,
    author = {Souza, Mateus S. and Barcelos, Ana M. and Nunes, Rafael C. and Akarsu, \"Ozg\"ur and Kumar, Suresh},
    title = "{Mapping the \ensuremath{\Lambda}$_{s}$CDM Scenario to f(T) Modified Gravity: Effects on Structure Growth Rate}",
    eprint = "2501.18031",
    archivePrefix = "arXiv",
    primaryClass = "astro-ph.CO",
    doi = "10.3390/universe11010002",
    journal = "Universe",
    volume = "11",
    number = "1",
    pages = "2",
    year = "2025"
}

@misc{Escamilla:2025imi,
    author = {Escamilla, Luis A. and Akarsu, \"Ozg\"ur and Di Valentino, Eleonora and \"Oz\"ulker, Emre and Vazquez, J. Alberto},
    title = "{Exploring the Growth-Index ($\gamma$) Tension with $\Lambda_{\rm s}$CDM}",
    eprint = "2503.12945",
    archivePrefix = "arXiv",
    primaryClass = "astro-ph.CO",
    month = "3",
    year = "2025"
}

@misc{Akarsu:2025dmj,
    author = {Akarsu, \"Ozg\"ur and Eingorn, Maxim and Perivolaropoulos, Leandros and Y\"ukselci, A. Emrah and Zhuk, Alexander},
    title = "{Dynamical dark energy with AdS-dS transitions vs. Baryon Acoustic Oscillations at $z =$ 2.3-2.4}",
    eprint = "2504.07299",
    archivePrefix = "arXiv",
    primaryClass = "astro-ph.CO",
    month = "4",
    year = "2025"
}

@misc{Akarsu:2025nns,
    author = {Akarsu, \"Ozg\"ur and Di Valentino, Eleonora and Vysko{\v c}il, Jir{\'i} and Yilmaz, Ezgi and Yukselci, A. Emrah and Zhuk, Alexander},
    title = "{Nonlinear Matter Power Spectrum from relativistic $N$-body Simulations: \ensuremath{\Lambda}$_{s}$CDM versus \ensuremath{\Lambda}CDM}",
    eprint = "2510.18741",
    archivePrefix = "arXiv",
    primaryClass = "astro-ph.CO",
    month = "10",
    year = "2025"
}

@article{Deffayet:2010qz,
    author = "Deffayet, Cedric and Pujolas, Oriol and Sawicki, Ignacy and Vikman, Alexander",
    title = "{Imperfect Dark Energy from Kinetic Gravity Braiding}",
    eprint = "1008.0048",
    archivePrefix = "arXiv",
    primaryClass = "hep-th",
    reportNumber = "CERN-PH-TH-2010-166",
    doi = "10.1088/1475-7516/2010/10/026",
    journal = "JCAP",
    volume = "10",
    pages = "026",
    year = "2010"
}

@article{Bengochea:2008gz,
    author = "Bengochea, Gabriel R. and Ferraro, Rafael",
    title = "{Dark torsion as the cosmic speed-up}",
    eprint = "0812.1205",
    archivePrefix = "arXiv",
    primaryClass = "astro-ph",
    doi = "10.1103/PhysRevD.79.124019",
    journal = "Phys. Rev. D",
    volume = "79",
    pages = "124019",
    year = "2009"
}

@article{Cai:2015emx,
    author = "Cai, Yi-Fu and Capozziello, Salvatore and De Laurentis, Mariafelicia and Saridakis, Emmanuel N.",
    title = "{f(T) teleparallel gravity and cosmology}",
    eprint = "1511.07586",
    archivePrefix = "arXiv",
    primaryClass = "gr-qc",
    doi = "10.1088/0034-4885/79/10/106901",
    journal = "Rept. Prog. Phys.",
    volume = "79",
    number = "10",
    pages = "106901",
    year = "2016"
}

@article{Pujolas:2011he,
    author = "Pujolas, Oriol and Sawicki, Ignacy and Vikman, Alexander",
    title = "{The Imperfect Fluid behind Kinetic Gravity Braiding}",
    eprint = "1103.5360",
    archivePrefix = "arXiv",
    primaryClass = "hep-th",
    reportNumber = "CERN-PH-TH-2010-210",
    doi = "10.1007/JHEP11(2011)156",
    journal = "JHEP",
    volume = "11",
    pages = "156",
    year = "2011"
}

@article{Kamionkowski:2014zda,
    author = "Kamionkowski, Marc and Pradler, Josef and Walker, Devin G. E.",
    title = "{Dark energy from the string axiverse}",
    eprint = "1409.0549",
    archivePrefix = "arXiv",
    primaryClass = "hep-ph",
    reportNumber = "SLAC-PUB-16085",
    doi = "10.1103/PhysRevLett.113.251302",
    journal = "Phys. Rev. Lett.",
    volume = "113",
    number = "25",
    pages = "251302",
    year = "2014"
}

@article{Emami:2016mrt,
    author = "Emami, Razieh and Grin, Daniel and Pradler, Josef and Raccanelli, Alvise and Kamionkowski, Marc",
    title = "{Cosmological tests of an axiverse-inspired quintessence field}",
    eprint = "1603.04851",
    archivePrefix = "arXiv",
    primaryClass = "astro-ph.CO",
    doi = "10.1103/PhysRevD.93.123005",
    journal = "Phys. Rev. D",
    volume = "93",
    number = "12",
    pages = "123005",
    year = "2016"
}

@article{Akarsu:2025ijk,
    author = {Akarsu, {\"O}zg{\"u}r and {\c{C}}am, Arman and Paraskevas, Evangelos A. and Perivolaropoulos, Leandros},
    title = "{Linear matter density perturbations in the {\ensuremath{\Lambda}}$_{s}$CDM model: Examining growth dynamics and addressing the S $_{8}$ tension}",
    eprint = "2502.20384",
    archivePrefix = "arXiv",
    primaryClass = "astro-ph.CO",
    doi = "10.1088/1475-7516/2025/08/089",
    journal = "JCAP",
    volume = "08",
    pages = "089",
    year = "2025"
}

@article{Nojiri:2010wj,
    author = "Nojiri, Shin'ichi and Odintsov, Sergei D.",
    title = "{Unified cosmic history in modified gravity: from F(R) theory to Lorentz non-invariant models}",
    eprint = "1011.0544",
    archivePrefix = "arXiv",
    primaryClass = "gr-qc",
    doi = "10.1016/j.physrep.2011.04.001",
    journal = "Phys. Rept.",
    volume = "505",
    pages = "59--144",
    year = "2011"
}

@article{Nojiri:2017ncd,
    author = "Nojiri, S. and Odintsov, S. D. and Oikonomou, V. K.",
    title = "{Modified Gravity Theories on a Nutshell: Inflation, Bounce and Late-time Evolution}",
    eprint = "1705.11098",
    archivePrefix = "arXiv",
    primaryClass = "gr-qc",
    reportNumber = "PHYS.REPT.-692-(2017)-1-104, Phys.Rept. 692 (2017) 1-104",
    doi = "10.1016/j.physrep.2017.06.001",
    journal = "Phys. Rept.",
    volume = "692",
    pages = "1--104",
    year = "2017"
}

@article{Bamba:2012cp,
    author = "Bamba, Kazuharu and Capozziello, Salvatore and Nojiri, Shin'ichi and Odintsov, Sergei D.",
    title = "{Dark energy cosmology: the equivalent description via different theoretical models and cosmography tests}",
    eprint = "1205.3421",
    archivePrefix = "arXiv",
    primaryClass = "gr-qc",
    doi = "10.1007/s10509-012-1181-8",
    journal = "Astrophys. Space Sci.",
    volume = "342",
    pages = "155--228",
    year = "2012"
}

@article{DiGennaro:2022ykp,
    author = "Di Gennaro, Sofia and Ong, Yen Chin",
    title = "{Sign Switching Dark Energy from a Running Barrow Entropy}",
    eprint = "2205.09311",
    archivePrefix = "arXiv",
    primaryClass = "gr-qc",
    doi = "10.3390/universe8100541",
    journal = "Universe",
    volume = "8",
    number = "10",
    pages = "541",
    year = "2022"
}

@article{Bouhmadi-Lopez:2025spo,
    author = "Bouhmadi-L{\'o}pez, Mariam and Ibarra-Uriondo, Be{\~n}at",
    title = "{Cosmological perturbations for smooth sign-switching dark energy models}",
    eprint = "2506.18992",
    archivePrefix = "arXiv",
    primaryClass = "gr-qc",
    doi = "10.1016/j.dark.2025.102129",
    journal = "Phys. Dark Univ.",
    volume = "50",
    pages = "102129",
    year = "2025"

}

@article{Bouhmadi-Lopez:2025ggl,
    author = "Bouhmadi-L{\'o}pez, Mariam and Ibarra-Uriondo, Be{\~n}at",
    title = "{Cosmographic analysis of sign-switching dark energy}",
    eprint = "2506.12139",
    archivePrefix = "arXiv",
    primaryClass = "gr-qc",
    doi = "10.1103/v1cl-pr54",
    journal = "Phys. Rev. D",
    volume = "112",
    number = "6",
    pages = "063559",
    year = "2025"
}

@misc{Ibarra-Uriondo:2026zbp,
    author = "Ibarra-Uriondo, Be{\~n}at and Bouhmadi-L{\'o}pez, Mariam",
    title = "{Sign-Switching Dark Energy: Smooth Transitions with Recent \textit{DESI DR2} Observations}",
    eprint = "2602.12347",
    archivePrefix = "arXiv",
    primaryClass = "astro-ph.CO",
    month = "2",
    year = "2026"
}

@misc{Akarsu:2026anp,
    author = {Akarsu, {\"O}zg{\"u}r and Caruana, Maria and Dialektopoulos, Konstantinos F. and Escamilla, Luis A. and Kahya, Emre O. and Levi Said, Jackson},
    title = "{Hints of sign-changing scalar field energy density and a transient acceleration phase at $z\sim 2$ from model-agnostic reconstructions}",
    eprint = "2602.08928",
    archivePrefix = "arXiv",
    primaryClass = "astro-ph.CO",
    month = "2",
    year = "2026"
}

\end{document}